\documentclass[twocolumn]{aa}  

\usepackage{siunitx}
\usepackage{textcomp, gensymb}
\usepackage{amsmath}
\usepackage{natbib}
\usepackage{graphicx}
\usepackage{xcolor}
\usepackage[normalem]{ulem}
\usepackage[varg]{txfonts}
\usepackage[]{hyperref}
\DeclareSIUnit\erg{erg}
\bibpunct{(}{)}{;}{a}{}{,} 
\makeatletter
\renewcommand*\aa@pageof{, page \thepage{} of \pageref*{LastPage}}
\makeatother
\titlerunning{A significant outburst in CL-AGN Mrk~1018}
\authorrunning{R. Brogan et al.}

\begin{document}

   \title{Still alive and kicking: A significant outburst in changing-look AGN Mrk~1018}

   \author{R.~Brogan\inst{1,}\inst{2}
   \and
   M.~Krumpe\inst{1}
   \and
   D.~Homan \inst{1}
   \and
   T.~Urrutia\inst{1}
   \and
   T.~Granzer\inst{1}
   \and
   B.~Husemann\inst{3}
   \and
   J.~Neumann\inst{4}
   \and
   M.~Gaspari\inst{5}
   \and
   S.~P.~Vaughan\inst{6,}\inst{7,}\inst{8,}\inst{9}
   \and
   S.~M.~Croom\inst{9,}\inst{10}
   \and
   F.~Combes\inst{11}
   \and
   M.~P\'erez Torres\inst{12,}\inst{13,}\inst{14}
   \and
   A.~Coil\inst{15}
   \and
   R.~McElroy\inst{16}
   \and
   N.~Winkel\inst{4}
   \and
   M.~Singha\inst{17}
   }
   \institute{Leibniz-Institut für Astrophysik (AIP), An der Sternwarte 16, 14482 Potsdam, Germany \\ 
   \email{rbrogan@aip.de}
   \and
   University of Potsdam, Institute of Physics and Astronomy, Karl-Liebknecht-Str. 24-25, 14476 Potsdam, Germany
   \and
   EUMETSAT, Eumetsat Allee 1, D-64295 Darmstadt, Germany
   \and
   Max-Planck-Institut für Astronomie, Königstuhl 17, D-69117 Heidelberg, Germany
   \and
   Department of Astrophysical Sciences, Princeton University, Princeton, NJ 08544, USA
   \and
   School of Mathematical and Physical Sciences, Macquarie University, NSW 2109, Australia
   \and
   Astronomy, Astrophysics and Astrophotonics Research Centre, Macquarie University, Sydney, NSW 2109
   \and
   Centre for Astrophysics and Supercomputing, School of Science, Swinburne University of Technology, Hawthorn, VIC 3122, Australia
   \and
   ARC Centre of Excellence for All Sky Astrophysics in 3 Dimensions (ASTRO 3D)
   \and
   Sydney Institute for Astronomy, School of Physics, University of Sydney, NSW 2006, Australia
   \and
   Observatoire de Paris, LERMA, Collège de France, CNRS, PSL University, Sorbonne University, 75014, Paris
   \and
   Instituto de Astrofísica de Andalucía (IAA-CSIC), Glorieta de la Astronomía s/n, 18008 Granada, Spain
   \and
   Facultad de Ciencias, Universidad de Zaragoza, Pedro Cerbuna 12, E-50009 Zaragoza, Spain
   \and
   School of Sciences, European University Cyprus, Diogenes street, Engomi, 1516 Nicosia, Cyprus 
   \and
   Department of Astronomy, University of California, San Diego, CA 92092, USA
   \and
   School of Mathematics and Physics, The University of Queensland, St Lucia, QLD 4072, Australia
   \and
   Department of Physics and Astronomy, University of Manitoba, Winnipeg, MB R3T 2N2, Canada
   }

   \date{Received 21 March 2023; accepted 19 July 2023}

 
  \abstract 
   {
   Changing-look active galactic nuclei (AGN) have been observed to change their optical spectral type. Mrk~1018 is particularly unique: first classified as a type 1.9 Seyfert galaxy, it transitioned to being a type 1 Seyfert galaxy a few years later before returning to its initial classification as a type 1.9 Seyfert galaxy after $\sim$30 years.
   } 
   {
   We present the results of a high-cadence optical monitoring programme that caught a major outburst in 2020. Due to sunblock, only the decline could be observed for $\sim$200 days. We studied X-ray, UV, optical, and infrared data before and after the outburst to investigate the responses of the AGN structures.
   }
   {
   We derived a $u'$-band light curve of the AGN contribution alone. The flux increased by a factor of $\sim$13. We confirmed this in other optical bands and determined the shape and speed of the decline in each waveband. The shapes of H$\beta$ and H$\alpha$ were analysed before and after the event. Two \textit{XMM-Newton} observations (X-ray and UV) from before and after the outburst were also exploited.
   }
   {
   The outburst is asymmetric, with a swifter rise than decline. The decline is best fit by a linear function, ruling out a tidal disruption event. The optical spectrum shows no change approximately eight months before and 17 months after. The UV flux is increased slightly after the outburst but the X-ray primary flux is unchanged. However, the 6.4~keV Iron line has doubled in strength. Infrared data taken 13 days after the observed optical peak already show an increased emission level as well. 
   }
   {
   Calculating the distance of the broad-line region and inner edge of the torus from the supermassive black hole can explain the multi-wavelength response to the outburst, in particular: i) the unchanged H$\beta$ and H$\alpha$ lines, ii) the unchanged primary X-ray spectral components, iii) the rapid and extended infrared response, as well as iv) the enhanced emission of the reflected 6.4~keV line. The outburst was due to a dramatic and short-lasting change in the intrinsic accretion rate. We discuss different models as potential causes.
   }

   \keywords{Galaxies: active - 
            Galaxies: nuclei - 
            Galaxies: Seyfert -
            X-rays: galaxies -
            Accretion, accretion disks - 
            Black hole physics}

   \maketitle
%

\section{Introduction} \label{intro}

Active galactic nuclei (AGN) are composed of supermassive black holes (SMBHs, $10^6$--$10^{10} M_{\odot}$) situated at the centre of a galaxy, which continuously accrete matter and are extremely luminous. Various structures are thought to be found around the central SMBH. These are the following, in order of distance from the central engine: an X-ray corona; an optical-UV accretion disc; fast- and slower-moving clouds of gas known as the broad- and narrow-line regions (BLR and NLR); and a dusty torus. Differing AGN types are classified based on the relative widths of their emission lines \citep{khachikian}. In the traditional picture, a type 1 AGN displays both broad and narrow optical emission lines, whereas a type 2 designation shows only narrow lines, since the line-of-sight view to the central engine is blocked by the torus. Between these two extremes lie a range of intermediate types, as outlined in \citet{osterbrock}. These are labelled Seyfert types 1.2, 1.5, 1.8, or 1.9 depending on the relative line widths; for example, a type 1.9 only has a broad H$\alpha$ and no H$\beta$ line, while a type 1.8 will have weak broad components in the H$\beta$ lines too. However, repeated spectral observations of AGN have shown that these objects are capable of switching between spectral types, often accompanied by a significant change in luminosity \citep{tohline}. These are known as changing-look AGN (CL-AGN). In recent years a large number of CL-AGN have been found through repeated spectral observations from large-scale spectroscopic surveys \citep{macleod2016, macleod2019, yang2018}.

The changing-look transition can be on timescales of a few years, which is much shorter than expected from the traditional model of a standard thin accretion disc \citep{lawrence}. There are several mechanisms that may be responsible for this behaviour, for example instabilities in the accretion disc \citep{review} or a tidal disruption event (TDE), that is when a star is consumed by the central black hole \citep{tde_rees, tde_phinney}. Some of these scenarios can be tested observationally, for example absorption is evident from the X-ray spectrum and brightening due to a TDE being thought to decrease with a characteristic $t^{-5/3}$ power law. Chaotic cold accretion (CCA; \citealt{gaspari13,gaspari17_cca}; see \citealt{gaspari20} for a review) is another proposed channel to explain AGN variability. In the CCA framework, multi-phase clouds condense out of the diffuse gaseous halo (galactic or intergalactic) and rain onto the central SMBH. At scales of $r\sim1$\,-\,100\,pc, such clouds collide inelastically, can cancel their angular momentum, fall onto the SMBH, and induce a rapid boost in the intrinsic AGN accretion rate. Such chaotic `raining' clouds can thus drive rapid changes in the light curves, up to several orders of magnitude (e.g. \citealt{maccagni21,mckinley22,olivares22}).

Mrk 1018 is a galaxy and AGN system with a redshift of 0.043 \citep{veron-cetty}. It is a post-merger remnant: the host galaxy is clearly irregular, with a tidal tail \citep{mcelroy}. It was first classified as a Seyfert 1.9 galaxy by a spectrum taken in 1979 and published in \citet{osterbrock}, with a broader H$\alpha$ than expected in a classical Seyfert 2 galaxy. The changing-look nature of the AGN was then revealed via a 1984 spectrum published in \citet{cohen}. The optical emission lines, including H$\beta$, were found to have broadened and the luminosity had significantly increased, reclassifying Mrk~1018 as a Seyfert 1 galaxy. The AGN remained in this bright state for approximately 30 years before a serendipitous Multi Unit Spectroscopic Explorer (MUSE) observation that found the AGN mid-transition -- in the process of returning to its former type 1.9 state \citep{mcelroy}. This means that Mrk~1018 has undergone a full state transition from a type 1.9 to a type 1 Seyfert galaxy and back to a type 1.9 Seyfert galaxy again, which cements Mrk~1018 as part of the group of extremely rare CL-AGN objects with multiple transitions. Similar AGN include the following: NGC~1566, which has experienced several events which caused the spectrum to range from a type 1.2 Seyfert galaxy to a type 1.9 Seyfert galaxy \citep{ngc1566}; and NGC~4151, which was first classified as a type 1.5 Seyfert galaxy \citep{ngc4151_osterbrock} and was observed to change to a type 1.8 Seyfert galaxy during two minimum states in 2001 and 2005 \citep{ngc4151_shapo}. Mrk~1018 is also one of the few CL-AGN to be observed at the time of transition (other examples are 1ES~1927+654, \citealt{trakhtenbrot} and Mrk~590, \citealt{denney_14}). This is a major bonus that gives valuable insight into AGN accretion physics and the physical origin of the CL-AGN phenomenon, since we can track the effects on the AGN structures as they are happening.

\citet{mrk1018_mass} estimated the mass of the central black hole to be $\log(M_{\textrm{BH}}/M_{\sun})=8.15$ with an uncertainty of 0.4 dex, using the virial method (e.g. \citealt{woo_urry_02}), and following the formula described in \citet{mcgill_08}. \citet{mcelroy} inferred a value of $\log(M_{\textrm{BH}}/M_{\sun})=7.9$ from an SDSS spectrum in 2009 following the formula described in \citet{woo_15}, which is within the error margins of the 
value from \citet{mrk1018_mass}. The same calculation in \citet{mcelroy} with the 
faint-phase MUSE spectrum from 2015 returns a value of $\log(M_{\textrm{BH}}/M_{\sun})=7.4$, most probably because the BLR was not in equilibrium or due to a change in the virial factor. Thus, we use the most recent value from a non-transitional period in this paper, that is $\log(M_{\textrm{BH}}/M_{\sun})=7.9$. 

Due to the fact that the AGN remained at a relatively stable high luminosity output for the previous 30 years, a TDE is excluded as the cause. X-ray follow-up observations also show that the change in spectral type and luminosity is not due to obscuration along the line of sight as a change in neutral hydrogen absorption ($N_{\rm H})$ is not detected \citep{husemann}. The UV-optical spectral energy distribution analysed during this time indicates a decrease in intrinsic accretion rate. The 2016 X-ray spectrum shows \citep{husemann} that a 6.4~keV Iron line is visible. This is thought to be a fluorescence line from the K-shell of Iron, produced at parsec-scale distances where the continuum radiation is reprocessed by circumnuclear material \citep{fe_origin}. \citet{lamassa} performs a detailed analysis of X-ray spectra from 2010 and 2016 and demonstrates that the Fe-line had not yet responded to the decrease in flux causing the shutdown. This time lag gives an indication of the distance of the X-ray reflecting gas from the accretion disc. 

As a galactic merger remnant it is plausible that a SMBH binary system (BBH) is hidden at the heart of the galaxy. This idea has been brought up before -- \citet{krumpe} state that the continuing significant short term variability of the AGN brightness after the type change could be due to a BBH. The gravitational interactions in a system such as this would affect the accretion rate periodically. \citet{Kim_2018} suggest the alternative scenario of a recoiling SMBH. In the recoil scenario, a galactic merger results in a central SMBH (the product of the two SMBHs of the respective galaxies) which is initially displaced by the kick velocity from the merger and oscillates in the galactic centre.

After catching Mrk~1018's second observed state change from a type 1 back to a type 1.9 Seyfert galaxy \citep{mcelroy, husemann, krumpe}, we continue to monitor the AGN in the optical for any further changes. We use the 1.2~m STELLA Observatory located on Tenerife, Spain \citep{stella_main, stella_plus}, which we have access to through the Leibniz Institute for Astrophysics Potsdam (AIP). A small outburst in 2017 is discussed in \citet{krumpe}, which is partially unobservable due to sunblock. Before Mrk~1018 becomes unobservable, a rise of $\sim$0.4 mag above the low state is observed. This increase is also mirrored in the UV and X-ray, which increase by a factor of $\sim$1.5 and $\sim$1.9 respectively.

The focus of this paper is to investigate the potential impact of the most significant outburst detected since Mrk~1018 returned to a Seyfert type 1.9. The outburst happens around mid-2020 and by itself would have caused Mrk~1018 to be flagged in a photometric search for CL-AGN. It is also partially obscured by the sunblock period and lasts only 200--300 days. A future paper (Brogan~et al.,~in prep.) will analyse the long-term multi-wavelength behaviour of Mrk~1018. This will include the global CL-transition of Mrk~1018 a few years ago as well as the outburst of 2020. In this paper we analyse multi-wavelength data taken before and after the optical outburst to probe the response of the various AGN structures. For this reason the sections on each type of data are split into the latest available observations before the outburst and the earliest available afterwards. Our data set includes X-ray and UV data, optical (photometric and spectroscopic) data, and IR data. 

Throughout the paper we use the wording `Mrk~1018' to refer to the entire system of AGN and host galaxy. When referring to the Mrk~1018 host, we specify the host galaxy and similarly when referring to the AGN. In Sect.~\ref{sec2}, we outline the multi-wavelength data we use. We present the details on how to create an so-called AGN-only optical light curve of Mrk~1018 based on STELLA data in Sect.~\ref{sec3}. We also analyse all multi-wavelength data to investigate any structural changes in this section. In Sect.~\ref{sec4}, we present our results and in Sect.~\ref{sec5} discuss their implications before concluding in Sect.~\ref{sec6}.

When we derive luminosities, we use the cosmological parameters $H_0 = 68$ km s$^{-1}$ Mpc$^{-1}$, $\Omega_m = 0.3$, and $\Omega_{\Lambda} = 0.7$ \citep{cosmo_params}. Using Mrk~1018's redshift ($z=0.043$), we infer a physical scale of 0.873 $\si{kpc}$ per arcsecond. The magnitude system used is that of SDSS, which is approximately similar to AB magnitudes. The photometric zero point in the SDSS $u$-band is +0.04 mag relative to the AB system \citep{sdss_mags}. Uncertainties are $1\sigma$ (68.3\% for one parameter) confidence intervals unless otherwise stated.

\section{Data} \label{sec2}

\subsection{Optical photometry}

\subsubsection{STELLA monitoring programme}

The STELLA telescopes are 1.2~m in diameter and this programme uses the Wide Field STELLA Imaging Photometer (WiFSIP) on the telescope STELLA-I \citep{stella_main, stella_plus}. STELLA operates differently than a traditional human-operated telescope as the available time is not apportioned to individual observers and objects. Rather, targets can be accessed at any point in time and a scheduling algorithm is used to maximise efficiency. Moreover, STELLA can react automatically to weather conditions and meteorological parameters.

\begin{figure}
   \centering
   \resizebox{\hsize}{!}{\includegraphics{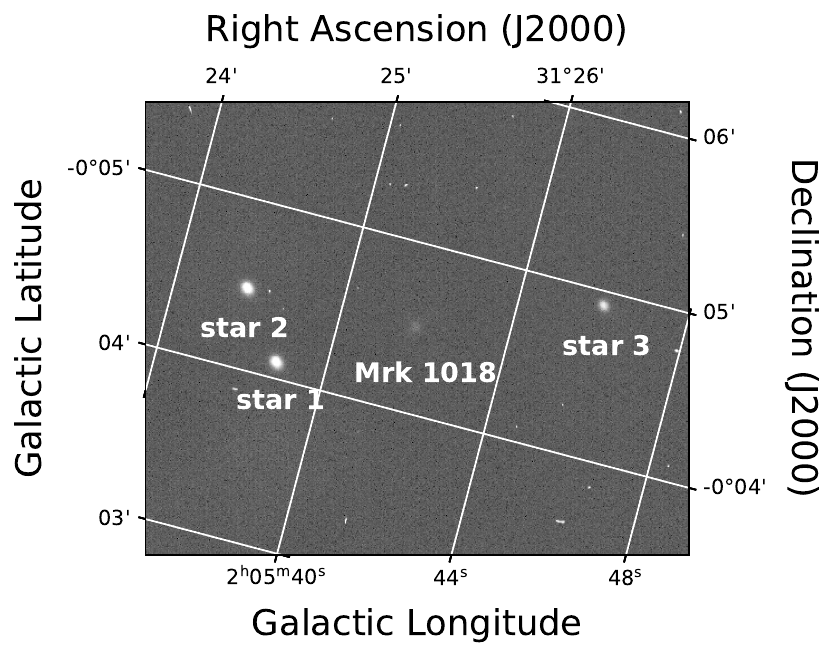}}
      \caption{Typical $u'$-band STELLA exposure of Mrk~1018 (at the centre of the image) and three reference stars. The galactic and ICRS coordinates are shown. The two stars on the left hand side are bright, and can be used to locate Mrk~1018 and the reference star on the right. The reference star on the right-hand side is used for relative photometry as it is unlikely to be saturated and has a similar $u'$-$g'$ colour to Mrk~1018.}
         \label{stellaexp}
   \end{figure}

The monitoring programme began on 22 October 2019. Our programme loosely follows a biweekly cadence and each observation consists of three images with exposure times ranging from 600--4800 seconds, depending on the visibility of Mrk~1018. An example of one of the images can be seen in Fig.~\ref{stellaexp}. The Sloan $u'$-band filter is chosen as the AGN is stronger in this filter in comparison to its host galaxy than in other optical filters. As is often the case with lower luminosity Seyfert galaxies, the light contribution of the host galaxy is significant in Mrk~1018. Each $u'$-band exposure covers an area of approximately 22' by 22'. The observations are corrected for bias and flat fielded, as described in \citet{weber_16}.

\subsubsection{VIMOS and GMOS images}
Through our collaboration with the Close AGN Reference Survey (CARS; \citealt{cars, carsb}), we have access to an additional 20 images of Mrk~1018 taken by the European Southern Observatory's (ESO) VIsible MultiObject Spectrograph (VIMOS; \citealt{vimos}) with exposure times of 300~s each and 21 images obtained by GMOS (the Gemini Multi-Object Spectrograph; \citealt{gmosn, gmoss}) with exposure times of 100~s each. 
All these data were taken in the time period from July 2017 to January 2019, which is after the AGN transformed from a type 1 Seyfert galaxy to a type 1.9, therefore during a faint phase, but before the new outburst outlined in this paper.

\subsection{Infrared photometry}

We use public archival data from the Wide-field Infrared Survey Explorer (\textit{WISE}; \citealt{wise}), which surveys the entire sky in infrared (IR), to investigate the IR response to the optical outburst. Infrared AGN emission is usually thought to originate in the torus, dominated by reprocessing of higher frequency (shorter wavelength) photons by the dust located there. This component is far enough away from the central engine that temperatures are lower and more dust is able to form. The original \textit{WISE} mission ended in 2010. In 2013 the survey was relaunched and a new round of observations with the \textit{WISE} satellite were undertaken as \textit{NEOWISE} \citep{neowise}. \textit{NEOWISE} observations are obtained in two wavebands: W1, centred at 3.4 $\mu$m and W2, centred at 4.6 $\mu$m. We note that the two filters used by \textit{NEOWISE} do not coincide with the peak of a black body of several 100 K as expected from the torus. However, they do provide important information about the overall strength of the IR output. The data and images are downloaded from the NASA/IPAC Infrared Science Archive\footnote{https://irsa.ipac.caltech.edu/Missions/wise.html}. These data have already been run through the \textit{NEOWISE} processing pipeline which firstly performs instrumental calibrations, detects sources in the images, calibrates for photometry and astrometry, and flags contaminating artefacts\footnote{https://wise2.ipac.caltech.edu/docs/release/neowise/expsup/sec4\_1.html}. We create a python script using the IRSA API (application programming interface), that downloads data on \textit{WISE} sources based on input co-ordinates and a user-defined radius. The source fluxes and magnitudes are then re-binned so that all entries are combined for each six-month \textit{WISE} epoch.

\subsection{Optical spectroscopy}

Due to the COVID pandemic and the global shutdown of almost all observing facilities, no optical spectra could be obtained during the outburst in 2020. Therefore, we analyse optical spectra before and after (as close in time as possible to the outburst) to investigate if any changes are seen in the BLR.

\subsubsection{Before the outburst}

The closest spectral observation pre-outburst was taken on 26 October 2019; approximately eight months before the observed peak of the $u'$-band outburst in 2020. This was taken with the FOcal Reducer/low dispersion Spectrograph 2 (FORS2; \citep{fors2}) on ESO's Very Large Telescope (VLT UT1) and details of the observation are found in \citet{hutsemekers}. The spectrum used in this paper is a combination of three observations taken on the same night using the grism 300V, with a wavelength range of 3300--6600~\AA. As this is a spectropolarimetric observation, there are observations available with different polarisation, as indicated by the Stokes parameters \citep{stokes}. This paper is not concerned with the polarisation of the spectrum, so we only used the unpolarised observations with Stokes parameter I. The raw data are downloaded from the ESO archive\footnote{http://archive.eso.org/cms.html} and reduced with the standard ESO FORS2 pipeline, the details of which are available in the documentation\footnote{http://www.eso.org/sci/software/pipelines/}.

Through our collaboration with CARS, we also obtained a proprietary data cube taken with the Kilofibre Optical AAT Lenslet Array (KOALA; \citealt{koala}) on the Anglo-Australian Telescope (AAT). This was observed on 4 September 2019, less than two months before the VLT observation. We took the spectrum from the central spaxel in the image, oriented on the AGN. However, this observation is taken with the 580V grating, which only covers the range 3700--5800~\AA. Consequently, only H$\beta$ is covered. This spectrum shows that the H$\beta$ line profile did not vary significantly on time-scales of months in the pre-outburst phase.

\subsubsection{After the outburst}

The first post-outburst optical spectrum was obtained with the Large Binocular Telescope's Multi-Object Double Spectrographs (LBT's MODS; \citealt{lbt}) on 3 December 2021. The object was exposed for 2000 seconds in total, using a $1\arcsec$ slit and a grating covering the wavelength range 3200--10\,000~\AA. We corrected the raw data for flat, bias and bad pixels using the modsCCDred script\footnote{http://www.astronomy.ohio-state.edu/MODS/Software/modsCCDRed/}, then extracted the spectra using IRAF's apall package\footnote{https://iraf.net/} \citep{iraf80, iraf90}. We took the source spectrum at the central position plus five pixels either side and took the background spectrum 20–30 pixels from the central position on both sides so that the host galaxy was not included. We also did wavelength and flux calibrations using IRAF. Finally, we combined the observations from MODS1 and MODS2.

\subsection{X-ray and UV data}

The \textit{XMM-Newton} X-ray satellite observed Mrk~1018 17--18~months before and 7--8~months after the observed peak of the optical outburst. On 4 January 2019 \textit{XMM-Newton} (ObsID: 0821240301) obtained a total exposure of 68~ks. The second observation is from 4 February 2021 (ObsID: 0864350101) and has a total exposure of 65~ks. In both observations the PN and mos cameras used the full frame window mode and a medium filter. The optical monitor (OM) onboard \textit{XMM-Newton} obtained simultaneous UV data with the X-ray data. In both observations the OM utilised the UVM2 filter which is centred at $\sim$230~\AA. Sequential 5000~s UVM2 images are collected during the entire X-ray observations.

\section{Analysis} \label{sec3}

\subsection{STELLA light curve} \label{stella_light_curve}

\subsubsection{Aperture photometry}

   \begin{figure}[t]
   \centering
   \resizebox{\hsize}{!}{\includegraphics{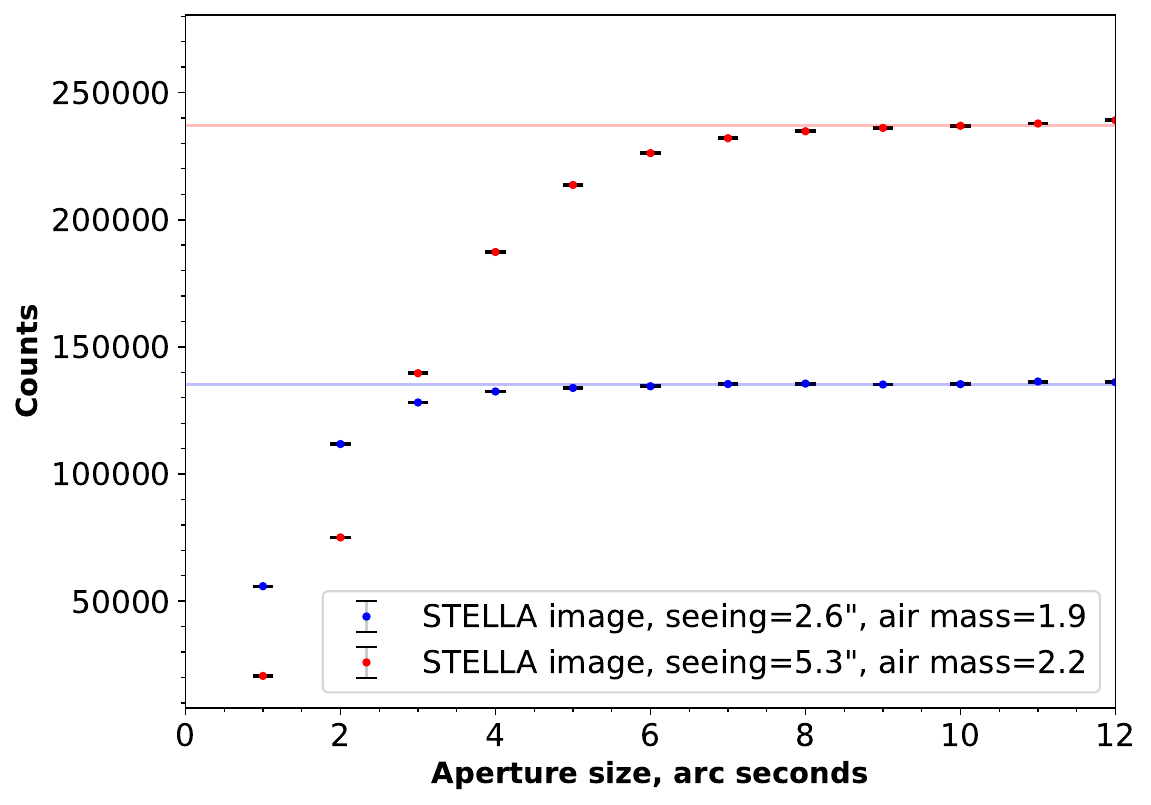}}
      \caption{Growth curve, i.e. the number of counts as a function of aperture radii. We measured the aperture photometry (integrated counts) for reference star three with increasing aperture size in $1\arcsec$ increments. The aperture photometry from two different STELLA images is shown. The pixel size of STELLA is 0.322\arcsec~per pixel. The uncertainties shown are the square roots of the counts. These images have seeing values on the low and high end of the range in our data set. The air mass is moderate and both have an exposure time of about 1200 seconds. The two horizontal lines indicate the count level at an aperture size of radius $10\arcsec$ -- the final aperture size chosen. The count number does not increase significantly above this value.
      }
         \label{growthcurve}
   \end{figure}

We visually screened all images (three per night) and discarded images affected by deformations (due to strong winds) and ghost patterns. This results in 20 nights that have at least one observation that survived our screening. All exposures are taken in the $u'$-band. We analysed the exposures photometrically using Astropy's PhotUtils package\footnote{https://photutils.readthedocs.io/en/stable/}. The background in STELLA images varies both spatially and in different exposures so a simple median is not sufficient. To compensate for the variable background, we used the PhotUtils Background2D\footnote{https://photutils.readthedocs.io/en/stable/background.htmlfunction}. We firstly divided the image into boxes of 35 by 40 pixels; we chose this size such that the box size is larger than the sources in the image, but small enough to capture the spatial variability of the STELLA background. We then performed sigma clipping with $\sigma=3$ in each area to remove sources from the background estimation in each box. This outputs an individual background map which can be subtracted from each image.

We used the PhotUtils source detection function to define a detection threshold as a number of standard deviations above the background count. We specified a detection threshold of 3-sigma. However, Mrk~1018 and one of the reference stars are not consistently bright enough to be identified in every STELLA image as sources, so we identified the two brighter reference stars first, allowing Mrk~1018 and the third reference star to be found using relative coordinates. A typical STELLA exposure of Mrk~1018 and the three reference stars is shown in Fig.~\ref{stellaexp}.

Once we located the sources and subtracted the background from the image, we performed aperture photometry to find the integrated counts within a circle. The aperture needs to be large enough to fully contain all the photons from each reference star, even at high seeing, and small enough that the signal is not lost in noise. To investigate this we chose two exposures with low air mass and long exposure times at either end of the range of seeing values. We used reference star three to plot a curve of growth, as shown in Fig. \ref{growthcurve}. The number of counts collected increases up to a (circular) aperture of radius $\sim$8\arcsec. We chose an aperture size of radius $10\arcsec$ for our analysis to compensate for any images with high seeing or slight deformities, for instance elongation. We then performed aperture photometry in each image taken during a night and then averaged the background-corrected counts over
all these images. We refer to the average value per night as an `observation'. These averages are weighted by the statistical error in the individual measurements.

   \begin{figure}[t]
   \centering
   \resizebox{\hsize}{!}{\includegraphics{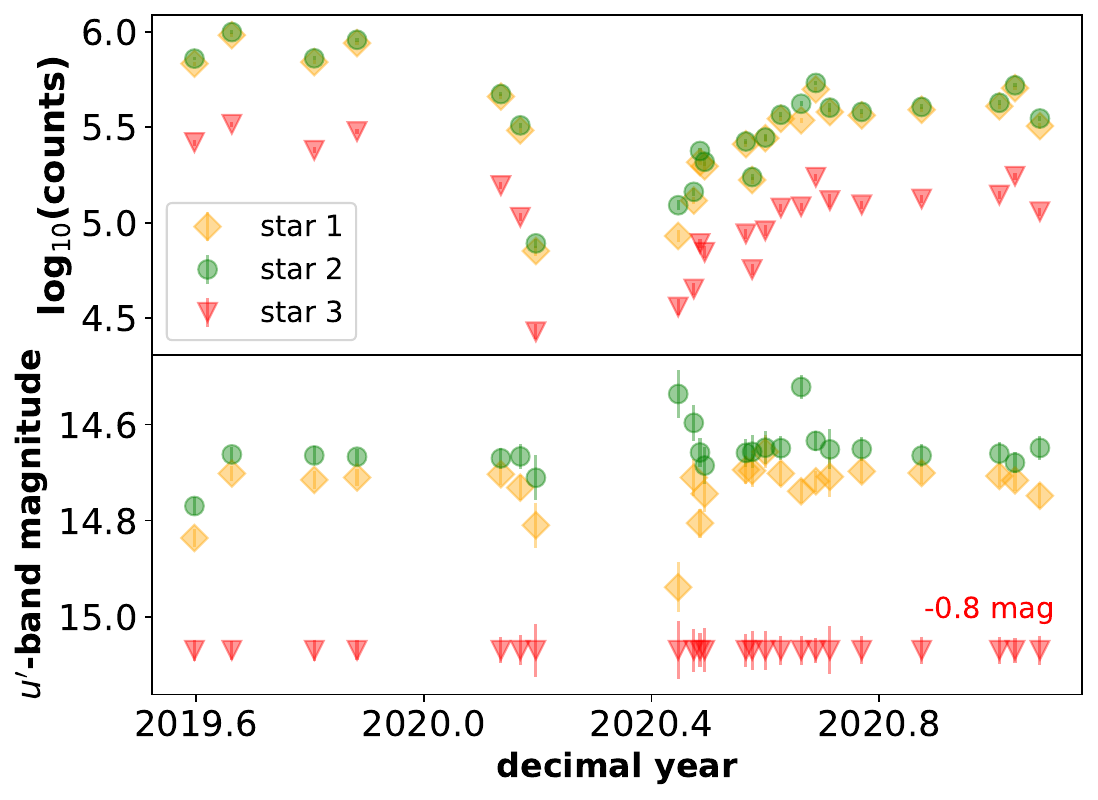}}
      \caption{Top: Average counts per $u'$-band observation (averaged over 1--3 background-corrected exposures) of the three reference stars. These are measured with a $10\arcsec$ radius aperture. There is a loss of $\sim$9\% in measured counts around the sunblock period (approximately 2020.2--2020.4) due to air masses as high as $\sim$5. Bottom: Three stellar light curves normalised to reference star three. These are obtained by differential photometry. To correct for the zero point, we added the SDSS magnitude for reference star three. We subtracted 0.8 magnitudes from reference star three in this plot for readability.}
         \label{starslc}
   \end{figure}

Using the reference stars around Mrk~1018, we explored potential instrumental effects. The upper panel of Fig.~\ref{starslc} shows the average integrated counts per observation for the STELLA data set. These light curves show that the three stars' light emissions vary in a similar manner and therefore we assumed that they remain at constant brightness throughout. Reference stars one and two have respectively bluer and redder $u'-g'$ colours than Mrk~1018 in the Sloan Digital Sky Survey (SDSS) database\footnote{https://skyserver.sdss.org/dr16/en/home.aspx} \citep{sdss}. The SDSS $u'$ and $g'$ magnitudes and $u'-g'$ colours are listed in Table~\ref{sloan} for Mrk 1018 and the reference stars. It is evident that reference star three is closest in colour to Mrk~1018 so we normalised to this one. The result of this normalisation is shown in the lower panel of Fig.~\ref{starslc}. Some differences between the stars can be seen around the sunblock period when the air mass is high (up to an air mass of 5.2; Mrk~1018 is low on the horizon at these times). This is due to their differing colours, as each star is attenuated differently by the Earth's atmosphere. Therefore, as reference star three is closest in colour to Mrk~1018, the attenuation effect should be the most comparable. We then created a light curve for Mrk~1018 by computing the differential magnitude with respect to reference star three and using the $u'$-band magnitude for reference star three as a zero point correction.

%
\begin{table}
\caption{SDSS magnitudes from DR16 for Mrk~1018 and the three reference stars. Reference star three is closest in colour to Mrk~1018, thus we used this star for flux normalisation.}       
\label{sloan}     
\centering                        
\begin{tabular}{c c c c}      
\hline           
Source & $u'$/[mag] & $g'$/[mag] & $u'-g'$/[mag] \\   
\hline                    
   Mrk~1018 & 15.64 & 14.30 & 1.34  \\     
   star 1 & 15.39 & 12.91 & 2.48 \\
   star 2 & 15.27 & 16.37 & -1.10  \\
   star 3 & 15.87 & 14.81 & 1.06  \\
\hline                                  
\end{tabular}
\end{table}

   \begin{figure}[t]
   \centering
   \resizebox{\hsize}{!}{\includegraphics{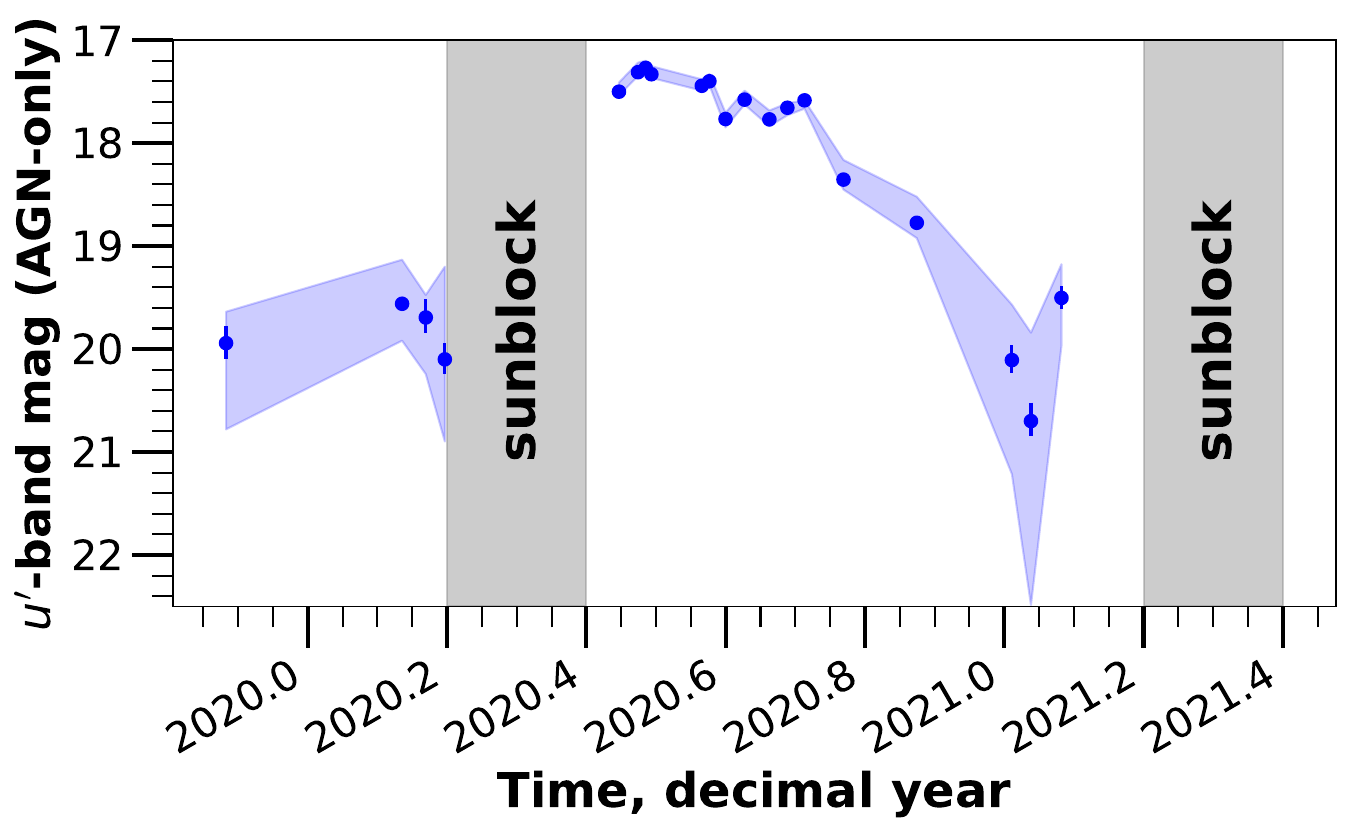}}
   \caption{Final host-subtracted $u'$-band light curve of the AGN in Mrk~1018. The blue points are the data points with statistical uncertainties shown by error bars and the blue shaded area represents the systematic uncertainty arising from the choice of image used to model the host galaxy component. This outburst is the most significant as yet observed during Mrk~1018's new type 1.9 state. Both sunblock periods are indicated in grey.}
              \label{stella_lc}%
    \end{figure}
%

\subsubsection{Host galaxy subtraction}

Aperture photometry measures the brightness of the host galaxy and AGN combined in the area chosen. The next stage of the reduction is to estimate how much the AGN alone contributes to the integrated counts by subtracting the host galaxy's contribution. The spatial resolution of STELLA is not sensitive enough to adequately decompose Mrk~1018 into its AGN and host components. Therefore, we used higher resolution images taken during the faint phase (VIMOS images in the time period 2016--2017) to model the host galaxy. We then produced a light curve of Mrk~1018's AGN without the host galaxy contribution, as shown in Fig. \ref{stella_lc}. In this paper we refer to this light curve as the AGN-only light curve. We also estimated the systematic uncertainties in the choice of high-resolution image for modelling, which are represented by the shaded region in Fig. \ref{stella_lc}. These systematic uncertainties dominate when the AGN is in its faint phase and have only a very minor contribution during the outburst in mid 2020. For a detailed description of the host galaxy subtraction procedure and calculation of the statistical and systematic uncertainties, see Appendix \ref{appendix:host_subtraction}.

The last data point in Fig.~7 is on 31 January 2021 because STELLA suffered a mechanical failure and it took several months to get back online. In the interest of demonstrating that the AGN was indeed in a semi-stable faint Seyfert type 1.9 phase before and after the 2020 outburst, we show the extended ATLAS optical light curve in Appendix \ref{appendix:big_atlas_lc}.

We also calculated the approximate fraction of the total energy output that the AGN in Mrk~1018 contributes over this period. We did this by taking the difference between Mrk~1018 (host galaxy plus AGN) and the AGN alone (see appendix~\ref{appendix:abfluxmag} for the flux-magnitude relation). The results show that the AGN consistently contributes less than 10\% (sometimes significantly less than 10\%) to the total flux in the faint phase and between 40\% and 50\% around the (observed) peak of the outburst in 2020.

\subsubsection{Light curve fitting}
\label{lightcurvefitting}

Using the AGN-only light curve, we modelled the outburst. This could reveal important insights into mechanisms that may have caused the outburst. Unfortunately, we missed the beginning of the outburst due to the sunblock period. Therefore, we focused on the shape of the decline. 

In our first fitting approach, we considered all data points from MJD 59023--59245. This includes points from the observed peak of the outburst until the last data point in Fig.~\ref{stella_lc}. We converted the magnitudes to fluxes by firstly adding the 0.04 mag offset between SDSS and AB mags. We then used the flux-magnitude relation for AB mags to calculate the fluxes in $\micro$Jy. We converted the magnitude uncertainties using the error propagation equation (see Appendix~\ref{appendix:abfluxmag} for all relations used).  We tested three simple models to find a best-fit: i) a linear function, ii) a parabolic function, and iii) a power-law function\footnote{Python fitting package lmfit: https://lmfit.github.io/lmfit-py/}. The latter was chosen in order to test if the light curve follows the characteristic shape of the decline of a TDE outburst. Canonically, this is described by a a power law with an exponent of $-5/3$ \citep{tde_rees, tde_phinney}. Although \citet{vanvelzen} show that this parameter can vary from $-2.0$ to $-0.8$, the mean power-law index of their sample is consistent with $p=-5/3$.

\begin{figure}[t]
   \centering
   \resizebox{\hsize}{!}{\includegraphics{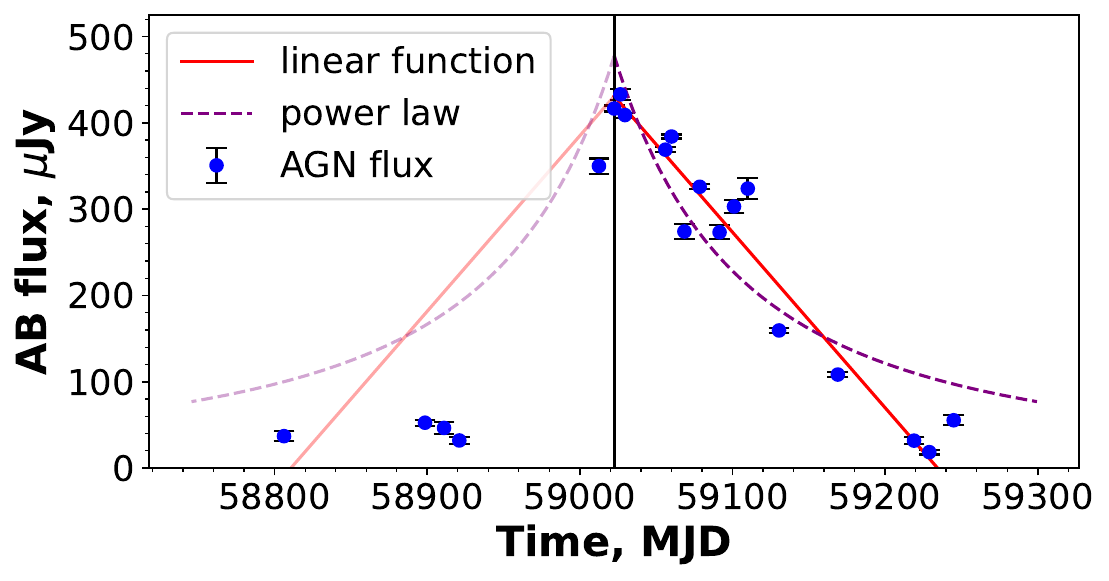}}
   \caption{Straight line and power law functions fitted to the decline of the 2020 outburst as observed in the $u'$-band. The black vertical line indicates where the fit starts. Only the data from this point onwards are fit as the rise of the outburst was hidden during sunblock. The fainter functions plotted on the left-hand side show the best fit functions mirrored for comparison with the data points before the outburst was observed. This implies an asymmetric outburst with a faster rise than decline.}
              \label{stella_fit}%
    \end{figure}

We note that the functions chosen are not intended to be a precise fit to the decline as the exact shape of the outburst cannot be determined due to sunblock. Fig.~\ref{stella_fit} shows the range of points considered and the linear and power-law functions plotted over the data. We also mirrored the functions around the fit initial point to compare the decline with the rise. This does not tally with the data before the outburst, indicating an asymmetric flare with a rise of maximum 100 days and decrease of minimum 200 days.

%
%
\begin{table*}
\renewcommand{\arraystretch}{2}
\caption{Best-fit (linear function, $y=mx+c$) parameters for the outburst decline using STELLA and ATLAS data. The main uncertainty is given by the fit and the uncertainty in brackets is the difference between fits when slightly adjusting the data ranges.}
 \label{lc_fit_params}
\centering                          
\begin{tabular}{c c c c}        
\hline    
\hline
Best-fit parameter & $u'$-band & $c$-band & $o$-band \\   
\hline
\hline
   m/[$\mu$Jy/day] & $-2.0\pm0.1~(0.1)$ & $-1.9\pm0.1~(0.2)$ & $-2.9\pm0.2~(0.1)$\\     
   c / [$10^5~\mu$Jy] & $1.2\pm0.6~(0.5)$  & $1.1\pm0.9~(0.1)$ & $1.7\pm0.1~(0.1)$\\
   \hline          
\end{tabular}
\end{table*}
The linear function $y=mx+c$ is the best fit, with $\chi^2/\textrm{d.o.f.}=68$, and the best-fit values are presented in Table~\ref{lc_fit_params}. The parabolic function $y = a(x-s)^2+b$ is unsuitable, with $\chi^2/\textrm{d.o.f.}=226$, as is the power law $y = a \times (\frac{x-x_0}{x_{\textrm{sc}}} + 1) ^{-5/3}$, with $\chi^2/\textrm{d.o.f.}=271$. We also tested several power-law indices in the range shown in \citet{vanvelzen}, however $\chi^2$/d.o.f. is similarly large for other exponents in this range.

In order to estimate the magnitude of the systematic uncertainties in these fits, we slightly adjusted the window that we used. Rather than using the entire range of the decline, we chose the brightest and faintest post-sunblock points and determined the difference between the best-fit values (of the different data sets). These systematic uncertainties are stated in Table~\ref{lc_fit_params} as the second set of uncertainties. A more in-depth estimation of the systematic uncertainties, using a variety of different data ranges, is outwith the scope of this paper. We find that the systematic uncertainty estimates are the same order of magnitude as the statistical uncertainties.

\subsection{ATLAS light curve}

\subsubsection{Robustness check} \label{robustness_check}

We used the forced optical photometry server at the Asteroid Terrestrial-impact Last Alert System, or ATLAS \citep{atlas}, to confirm the outburst as seen in our STELLA light curve. The forced photometry server\footnote{https://fallingstar-data.com/forcedphot/} \citep{shingles_atlasweb} allows the user to request photometry for any sky coordinates as far back as the beginning of the ATLAS project and create a light curve using AB magnitudes in two optical filters. These filters are named `cyan', with wavelength range of 4200--6500~\AA, and `orange', with a range of 5600--8200~\AA. In order to create light curves of variable objects, the pipeline uses so-called difference images. These images are created by subtracting each observation from a matched all-sky `reference' image, giving a resultant flux difference in $\micro$Jy. The forced photometry is done by fitting a point-spread function to nearby high-resolution stars and forcing this fit at the input coordinates. The outburst can also clearly be seen in this data set.

\begin{figure}[t]
   \centering
   \resizebox{\hsize}{!}{\includegraphics{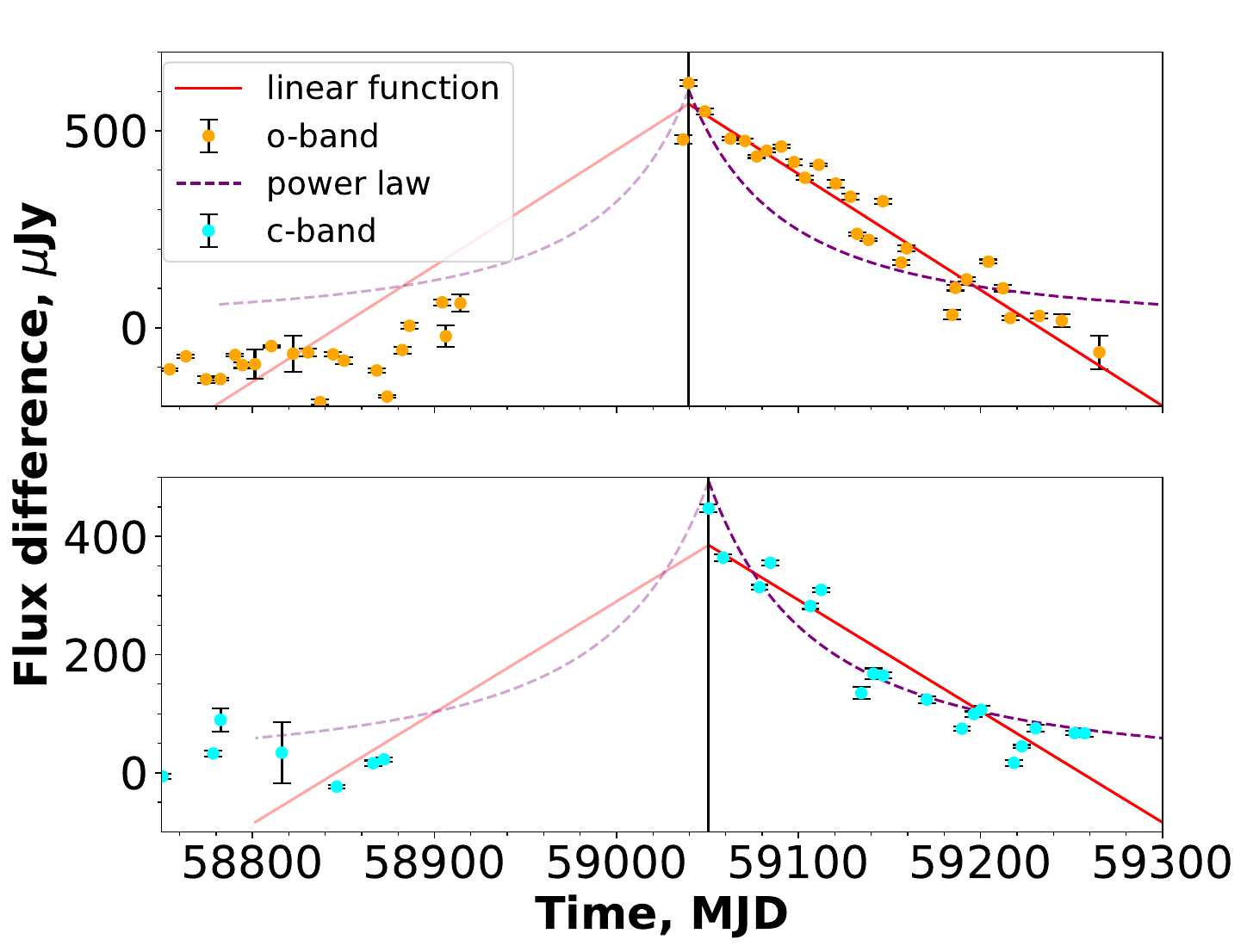}}
   \caption{Straight line and power-law functions fitted to the decline of the 2020 outburst as seen in the orange and cyan filters used by ATLAS. The orange and cyan data points represent observations in each filter respectively. The black vertical lines indicate where the fit starts. These are slightly different for each filter as the fit here was chosen to begin with the brightest data point in each data set. The functions are mirrored for comparison with the outburst rise.}
              \label{atlas_fit}%
    \end{figure}

Figure~\ref{atlas_fit} shows the light curve in the $o$- and $c$-bands as outputted by the ATLAS forced photometry server. This data set covers the same time period as the STELLA data. The observations are binned to a 7-day time-span using the script linked to on the ATLAS website\footnote{https://gist.github.com/thespacedoctor/86777fa5a9567b7939e8d84fd \ 8cf6a76}. This reduces scatter and makes the plot less crowded. The gaps before and after the outburst are during Mrk~1018's yearly sunblock period. Unlike the complex STELLA pipeline described in Sect.~\ref{degrading}, the host subtraction is performed implicitly by ATLAS with their so-called difference images. The host galaxy contribution and the flux of the AGN in the reference image is subtracted out with this method. Assuming that the host galaxy is constant and always in the aperture, this leaves only the relative amount by which the AGN brightened or dimmed. As a result, the ATLAS light curve shows a clear rise and decline of the AGN luminosity, confirming the outburst that was tracked with the STELLA data set.

\subsubsection{Light curve fitting}
\label{atlas_fit_sec}

We repeated the light curve fitting procedure, outlined in Sect.~\ref{lightcurvefitting}, for the ATLAS data. Results are shown in Table~\ref{lc_fit_params} and Fig.~\ref{atlas_fit}. For the $c$-band, the linear, parabolic and power-law functions have reduced $\chi^2$ statistics of 76, 176 and 109 respectively. For the $o$-band, the reduced $\chi^2$ statistics are 64, 146 and 127, respectively. Again, the linear function is the best fit for both the $c$- and $o$-bands.

We note that both ATLAS filter light curves look smoother then the STELLA light curve. We averaged the ATLAS data points over seven days to obtain a decent signal-to-noise quality, whereas the STELLA data points are single-night observations. Thus, naturally, STELLA data can show the intrinsic variability of the source on time-scales of a night, whereas the ATLAS data is smoothed out over a whole week.

   \begin{figure*}[t]
   \centering
   \includegraphics[width=17cm]{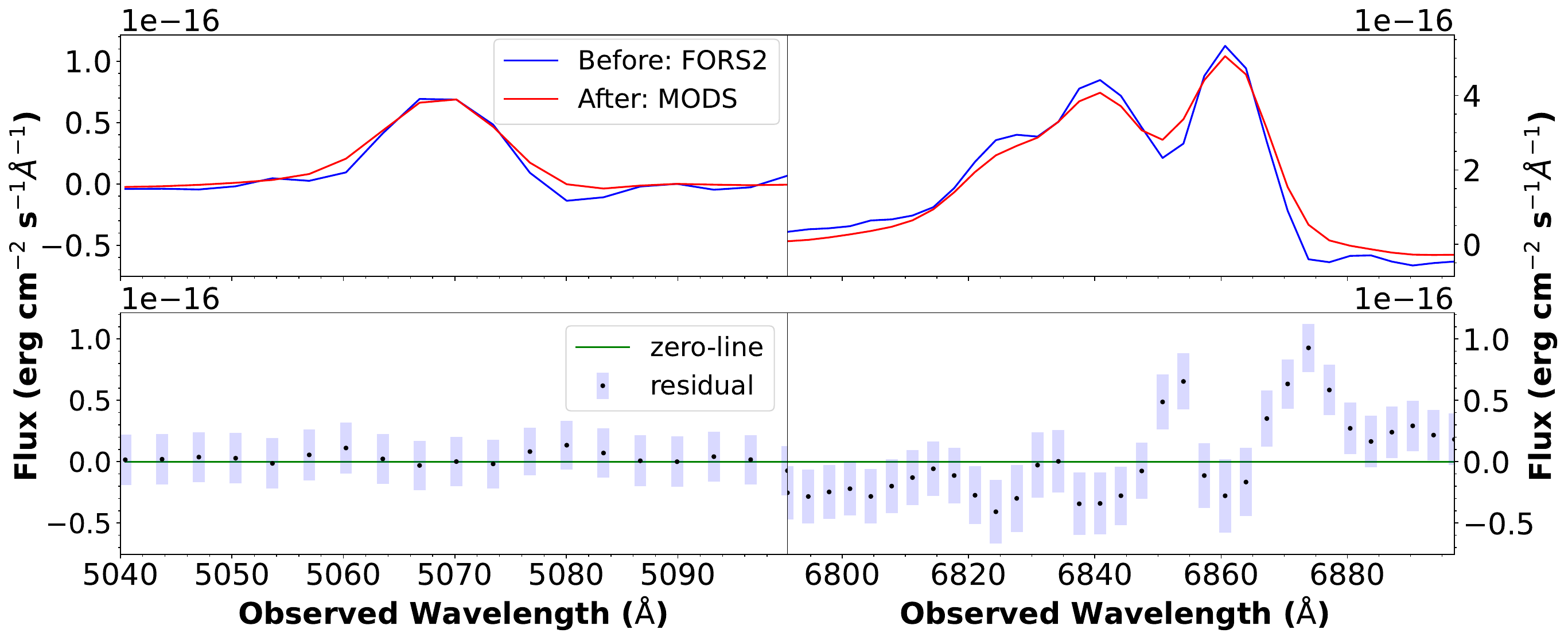}
   \caption{
   Comparison of spectral observations taken before and after the 2020 outburst, focusing on the H$\beta$ and H$\alpha$ emission lines. An appropriate value is subtracted to match continuum levels. Top left: Close up of the H$\beta$ wavelength range. Spectra were scaled using the average flux of the narrow [OIII] lines. Bottom left: Residuals (MODS$-$FORS2) for H$\beta$. The combined uncertainties of the two spectra are shown as blue shaded bars for each spectral bin. Top right: A close up of H$\alpha$ and the NII doublet. Scaling was done by comparing the average flux of the narrow [SII] lines. Bottom right: Residuals and combined uncertainties for H$\alpha$.}
    \label{optspec}
    \end{figure*}

\subsection{Comparison of spectra before and after the outburst}

In our spectral comparison we focused on the Balmer emission lines. We therefore only compared the spectra in two wavelength ranges in the vicinity of H$\alpha$ and H$\beta$ to determine whether any changes appear in these lines. We normalised each wavelength section using the integrated flux of nearest unblended narrow lines: the [OIII] doublet for H$\beta$ and the [SII] doublet for H$\alpha$. The NLR is much further away from the central engine (of the order of $10^2$ parsecs) than the BLR (10--100 light days away). Therefore any accretion changes should not affect the narrow lines over the short time-span these observations cover, but rather take decades to centuries to respond.

We started by fitting the H$\beta$ line. Firstly, we subtracted a constant value over a small wavelength range of $\sim$100~\AA to remove the continuum below the emission lines. We did not subtract the host galaxy contribution as we are merely comparing the two Balmer lines. Next, we fitted the [OIII] lines with single Gaussian functions in the individual spectra. We then determined the ratio of the fitted integrated fluxes between the spectra and used the average of the [OIII] integrated flux ratios to scale the spectra. We wavelength-matched the spectra using the fitted central wavelengths of the redder [OIII] line. We then re-binned the MODS spectrum to the FORS2 binning so that each data point is at the same wavelength as its counterpart. Due the difference in spatial resolution of the spectrographs -- MODS has a spectral resolution of 0.5~\AA/pixel and FORS2 of 1.68~\AA/pixel -- the MODS spectrum was convolved with a Gaussian with a FWHM of 3.36 to match the FORS2 resolution. The next step was to fit both H$\beta$ lines with a single Gaussian each. Since the single-Gaussian line-profile model leaves no significant residuals compared to the data, we refrained from applying more complex line models. The results for the best-fit models agree within their combined 2$\sigma$ uncertainties and are shown in Table~\ref{line_fits}. The left panels in Fig.~\ref{optspec} show the residuals between the two observed spectra.

We also scaled and fitted a KOALA spectrum from 2019 covering H$\beta$, but not H$\alpha$, to the MODS spectrum. Again, the MODS spectrum had to be degraded to the other spectrum's resolution -- KOALA has a resolution of 1.03~\AA/pixel. All line-fit parameters agree within their combined 2$\sigma$ uncertainties for the KOALA (before the outburst) and MODS spectrum (after the outburst), as shown in Table~\ref{line_fits}. This is consistent with our finding of the H$\beta$ line parameters based on the FORS2 and MODS spectra (before and after the outburst).

Following the same approach, we fitted the H$\alpha$ lines in the spectra before and after the outburst. The [SII] lines are used to scale the FORS2 and MODS spectra. Since H$\alpha$ is blended with the NII doublet (blue- and red-wards of H$\alpha$), we fitted three Gaussian line profiles to the line complex. The H$\alpha$ parameter values are shown in Table~\ref{line_fits} and their values agree within the fitting uncertainties before and after the outburst. We show the H$\alpha$ line comparison between the spectra in Fig.~\ref{optspec} (right).

%
\begin{table}[t]
\renewcommand{\arraystretch}{1.5}
\caption{Results for the H$\beta$ and H$\alpha$ line fittings before and after the outburst. The MODS spectral fitting parameters (after the outburst) are given after degrading the spectrum to the lower spectral resolution data of KOALA and FORS2 respectively. This allows for a meaningful comparison of the fitting parameters before and after the outburst, but not for a comparison between the two spectra before the outburst (i.e. FORS2 to KOALA). The observed FWHM is given in units of \AA and the integrated flux in units of $10^{-14}$~\si{erg.cm^{-2}.s^{-1}}.}
\label{line_fits}     
\centering                          
\begin{tabular}{c c c}        
\hline \hline
 FORS2 resolution \\
 \hline \hline
H$\beta$ & Before & After \\
 \hline
 Date (YYYY-MM-DD) & 2019-10-26 & 2021-12-03 \\
 Instrument & FORS2 & MODS \\
 Observed FWHM & 12.7 $\pm$ 1.3 & 12.6 $\pm$ 0.6 \\
 Integrated flux &  0.118 $\pm$ 0.011 & 0.100 $\pm$ 0.004 \\
 \hline \hline
KOALA resolution \\
 \hline \hline
 Date (YYYY-MM-DD) & 2019-09-04 & 2021-12-03 \\
 Instrument & KOALA & MODS \\
 Observed FWHM & 6.2 $\pm$ 0.6 & 6.4 $\pm$ 0.2 \\
 Integrated flux & 0.122 $\pm$ 0.011 & 0.103 $\pm$ 0.002 \\
\hline \\
 \hline \hline
FORS2 resolution \\
\hline \hline
H$\alpha$ & Before & After \\
 \hline
 Date (YYYY-MM-DD) & 2019-10-26 & 2021-12-03 \\
 Instrument & FORS2 & MODS \\
 Observed FWHM & 13.8 $\pm$ 2.1 & 14.1 $\pm$ 1.0 \\
 Integrated flux & 0.5 $\pm$ 0.2 & 0.4 $\pm$ 0.1 \\
 \hline
\end{tabular}
\end{table}

\begin{figure}[t]
   \hbox{\hspace*{-0.80cm}
   \includegraphics[width=9.5cm,height=6.5cm]{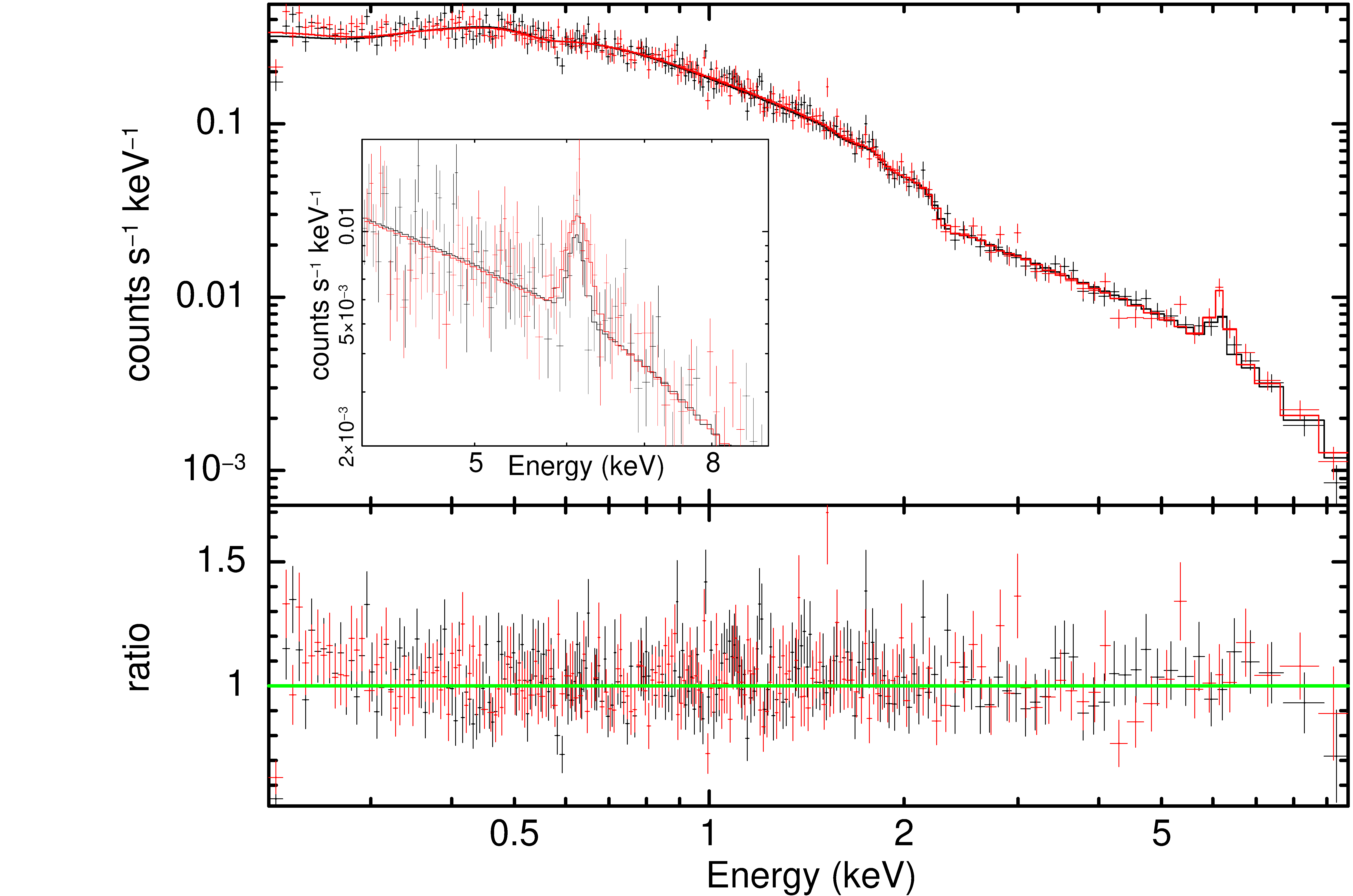}}
   \caption{Comparison between the PN (patter 0) spectra before (black) and after (red) the outburst. Each data set is shown with its corresponding best fit. The inset shows a zoom on the Fe line. For the sake of readability, we do not show the MOS1 and MOS2 spectra in this plot.}
              \label{xmm}%
    \end{figure}

%

\subsection{XMM-Newton spectra}

We extracted X-ray spectra before and after the outburst using the {\tt SAS} package (20.0.0) and {\tt HEASOFT} (v6.29), excluding observation times subject to an increased high energy background level. This left us 49~ks of good integration time in PN and 58~ks in each individual MOS for the 2019 spectrum, while the 2021 spectrum yielded 44 ks and 56 ks cleaned data for PN and each individual MOS, respectively. Pile-up does not affect these observations. We then created standard source and background spectra in PN, MOS1, and MOS2. We produced two spectra for PN: one only contained the events with pattern equal to zero (single events) and the other with patterns 1--4 (double). By not merging single and double events we increased the energy resolution in the single event spectrum. We grouped the spectra with a minimum of 20 counts per bin. Figure~\ref{xmm} shows the resultant spectra. We fitted the X-ray spectra using {\tt XSPEC} version 12.13.0b \citep{arnaud} in the 0.2--10 keV energy range, except for the patter 1--4 that we fitted in the 0.5--10 keV range. We used the cosmic abundances of \citet{wilms} and the photoelectric absorption cross section provided by \citet{verner}. 

Firstly, we fitted a model consisting of a power law, a narrow Gaussian line profile with an initial rest-frame energy of 6.4 keV and line width $\sigma=0.1$ keV, as well as Galactic neutral hydrogen absorption ($N_{\rm HI,Gal}=2.43 \times10^{20}$ cm$^{-2}$, \citealt{kalberla}). The fit is acceptable ($\chi^2$/d.o.f.=1458/1369=1.07), but above 5 keV the model underestimates the data. Thus, we included a second power-law component to improve the fit in the high energy range. This model leaves no obvious residuals when compared with the data. The fits yields $\chi^2$/d.o.f.=1396/1367=1.02. Allowing for an additional intrinsic absorption component does not improve the fit further. Thus, we consider the two power-law component plus Gaussian line profile our best-fit model. In Table~\ref{xray_fit}, we state the best-fit parameters and their uncertainties. As we only aimed to compare X-ray spectra before and after the outburst, we did not focus on finding the most likely physical model. We are only interested in finding a good model to search for obvious changes in both X-ray spectra. The Fe line is detected at a 6$\sigma$ level, with an equivalent width of 0.19 keV. The line energy is statistically consistent with a neutral Fe line at a rest-frame energy of 6.4 keV and the line is unresolved. 

We repeated the fitting procedure above for the \textit{XMM-Newton} spectrum after the 2020 outburst. We used a single power law component, a narrow Gaussian line profile, and Galactic absorption, which yields a reasonable fit ($\chi^2$/d.o.f.=1353/1338=1.01), but the data at energies above 6 keV show positive offsets compared to the model. Thus, we added a second power-law component and the fit improved ($\chi^2$/d.o.f.=1319/1336=0.99). After the outburst, the Fe line is also detected at a 6$\sigma$ level and the line energy is still consistent with a neutral Fe emission. However, the line width is consistent with being marginally resolved ($\sim$3$\sigma$) and has an equivalent width of 0.39 keV (twice as high as before the outburst). Therefore, although we cannot observe any changes between the primary X-ray flux of the two observations, the effect on the Fe line is still visible.
  
\begin{table}
\setlength{\tabcolsep}{4pt}
\renewcommand{\arraystretch}{2}
\caption{Best-fit X-ray parameters before and after the outburst, as well as their statistical differences.}    
\label{xray_fit}
\centering                       
\begin{tabular}{c c c c}
\hline           
Parameter & 2019 & 2020  & $\Delta[\sigma]$ \\
\hline

$\Gamma_1$ & $1.86_{-0.02}^{+0.02}$ & $1.83_{-0.02}^{+0.02}$ & 1.1 \\

$n_1$ / [10$^{-4}$ photons keV$^{-1}$ cm$^{-2}$ s$^{-1}$] & $2.46_{-0.06}^{+0.04}$ & $2.54_{-0.04}^{+0.03}$ & 1.4 \\

$\Gamma_2$ & $0.30_{-0.37}^{+0.27}$ & $-0.14_{-0.48}^{+0.44}$ & 0.8 \\

$n_2$ / [$10^{-4}$ photons keV$^{-1}$ cm$^{-2}$ s$^{-1}$] & $0.06_{-0.03}^{+0.05}$ & $0.02_{-0.01}^{+0.03}$ & 0.9 \\

$E_{Fe}$ / [keV] & $6.38_{-0.02}^{+0.02}$ & $6.39_{-0.02}^{+0.02}$ & 0.4 \\

$\sigma_{Fe}$ / [keV] & $0.01_{-0.01}^{+0.05}$ & $0.10_{-0.03}^{+0.03}$ & 1.5 \\

$n_{Fe}$ / [$10^{-6}$ photons cm$^{-2}$ s$^{-1}$] & $2.20_{-0.36}^{+0.35}$ & $4.53_{-0.62}^{+0.65}$ & 3.3 \\

$f_{0.2-2~\textrm{keV}}$ / [$10^{-13}$~\si{erg.cm^{-2}.s^{-1}}] & $6.53_{-0.04}^{+0.04}$ & $6.61_{-0.04}^{+0.04}$ & 1.4 \\

$f_{2-10~\textrm{keV}}$ / [$10^{-12}$~\si{erg.cm^{-2}.s^{-1}}] & $1.02_{-0.03}^{+0.03}$ & $1.01_{-0.02}^{+0.02}$ & 0.3 \\

$L_{2-10~\textrm{keV}}$ / [$10^{42}$~\si{erg.s^{-1}}] & $4.03_{-0.12}^{+0.12}$ &  $4.08_{-0.08}^{+0.08}$ &
0.3 \\

$f_{0.2-10~\textrm{keV}}$ / [$10^{-12}$~\si{erg.cm^{-2}.s^{-1}}] & $1.67_{-0.02}^{+0.03}$ & $1.67_{-0.02}^{+0.02}$ & 0.0 \\ 

$L_{0.2-10~\textrm{keV}}$ / [$10^{42}$~\si{erg.s^{-1}}] & $6.66_{-0.14}^{+0.20}$ & $6.75_{-0.14}^{+0.14}$ & 0.4 \\
\hline                           
\end{tabular}
\tablefoot{Parameters are based on the best fit of a model. The model used was {\tt TBabs(zpowerlw + zTBabs(zpowerlw) + zgauss}. The redshift was frozen at 0.042 and neutral hydrogen was frozen at $2.43\times10^{20}$ cm$^{-2}$. We did not correct the observed fluxes for Galactic absorption whereas the rest-frame luminosities were corrected for Galactic absorption.}
\end{table}

\subsection{UV data}

We reduced the UVM2 OM data with the standard OM \texttt{SAS} commands \texttt{omfchain} and \texttt{omichain}. These commands automatically produce images of the individual exposures as well as magnitudes and fluxes of the detected sources in these images. We confirmed by visual inspection that Mrk 1018 and the three reference stars are always present and detected in the images. The OM data reduction also provides averaged magnitudes and fluxes (including their statistical uncertainties) for all four targets of interest in both \textit{XMM-Newton} observations. 

We then evaluated the systematic uncertainties, comparing the magnitudes of the reference stars before and after the 2020 outburst and assuming no changes in their UV output. We show the corresponding magnitudes in Table~\ref{uv}. Systematic offsets of up to 0.05 magnitudes are found between the reference stars. However, Mrk~1018 brightened by $\sim$0.6 mag (a factor of 1.8) from 2019--2021. This brightening is even visible in the individual images of both observing campaigns. Considering the combined statistical and systematic uncertainties, we detect the brightening with a confidence of 11$\sigma$. 

%
\begin{table}
\renewcommand{\arraystretch}{1.3}
\caption{UV magnitudes for Mrk~1018 and the three reference stars before and after the outburst. These measurements are obtained simultaneously with the X-ray data using \textit{XMM-Newton}.}
 \label{uv}
\centering                         
\begin{tabular}{c c c} 
\hline
 & 2019-01-04 & 2021-02-04 \\
\hline
Object & Magnitude & Magnitude \\   
\hline
ref. star 1 & $18.3\pm0.1$ & $18.3\pm0.1$ \\
ref. star 2 & $17.9\pm0.1$ & $17.9\pm0.1$ \\
ref. star 3 & $17.9\pm0.1$ & $17.9\pm0.1$ \\
Mrk~1018 & $19.4\pm0.1$ & $18.7\pm0.1$ \\
\hline
\end{tabular}
\end{table}

\section{Results} \label{sec4}

\subsection{Optical photometry}

We detected a strong outburst from the AGN within Mrk~1018 in 2020. This was observed with the STELLA monitoring programme set up after the recent changing-look transition. Figure \ref{stella_lc} clearly shows a significant increase in the $u'$-band immediately following the Mrk~1018 sunblock period. The magnitude difference between the data point immediately before sunblock and the brightest data point is $\sim$2.8 mags, corresponding to an increase in flux by a factor of the order of 13. After the outburst it returns to approximately the same level -- less than 10\% of Mrk~1018's optical output. The length of the outburst is difficult to quantify as the ignition period is obscured due to sunblock. The length of time between the last observation before sunblock and the first one after is approximately 100 days. Since the last observation before sunblock shows no increase but the first observation after sunblock already shows a decline, the outburst rise cannot have lasted longer than 100 days. However, the rise itself is completely unobserved. Following sunblock, the decline is observed over about 200 days. The AGN returns to its faint state luminosity in approximately 300 days after the last observation before sunblock. The outburst is asymmetric and the increase took at most half the time of the decline.

The forced photometry server from ATLAS also observes this outburst, confirming the STELLA observations. ATLAS uses optical filters at longer wavelengths than the STELLA monitoring programme, therefore the host galaxy contribution is more pronounced in these data. Whereas we took care to make a detailed approximation of the host galaxy in our images and subtract this from our photometry, the ATLAS server simply uses differential imaging. As the host galaxy flux remains relatively stable, this contribution disappears when subtracting the images and we are left with the variable component, namely the AGN. This is a faster but less robust method of discounting the host flux. If we take the difference between data points similar in time in the three optical filters used, the largest magnitude difference is in the reddest filter; the $o$-band. There is less of a difference between the two bluer filters; the $u'$- and $c$-band. The outburst seen here is also asymmetric, with a steeper rise than decline. Again the length of the ATLAS outburst is obscured by sun avoidance, but is several hundreds of days, similar to that of STELLA.

\subsection{Comparison of optical light curves}

\begin{figure}[t]
   \centering
   \resizebox{\hsize}{!}{\includegraphics{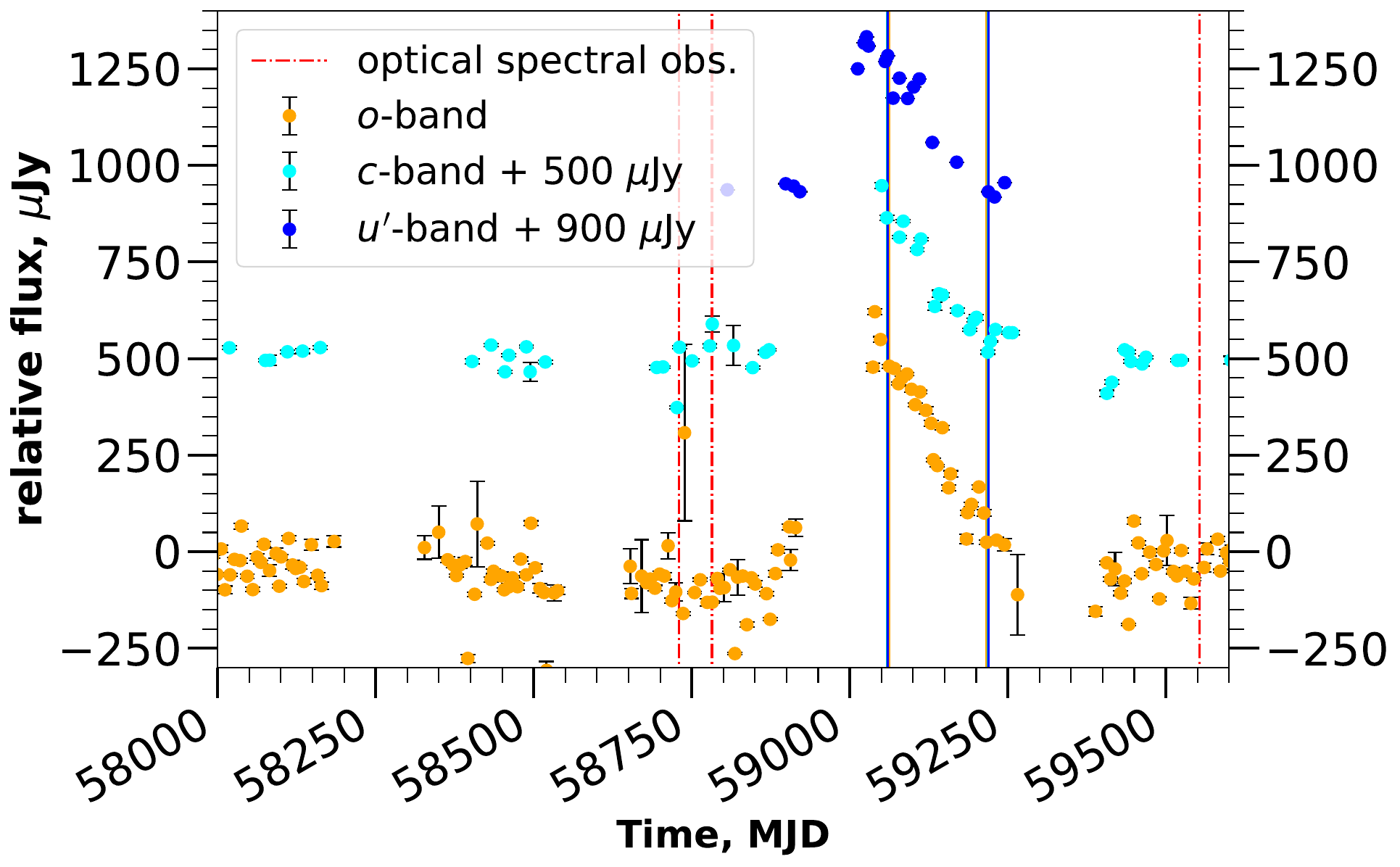}}
   \caption{Outburst as seen in the $u'$-band (STELLA), $c$-band (ATLAS) and $o$-band (ATLAS). The vertical lines with the same colours as the relevant data mark the points taken to calculate and compare the change in flux seen over the decline. The $u'$- and $c$-band data have been shifted up for ease of presentation. The dates of the optical spectral observations are shown by the red dot-dashed lines.}
              \label{del_mag_fig}%
    \end{figure}

Since we observed the decline in three sets of data, we could also compare the amplitude of the decline in the $u'$-, $c$- and $o$-filters, despite the different subtraction techniques\footnote{For the STELLA data we do a sophisticated host galaxy subtraction, while the ATLAS data are based on subtracting a reference image from all other images.}. To quote values that can be meaningfully compared, the time period over which the changes in the three bands are measured should be as similar as possible. We chose start and end points at two sufficiently distant times when data from all different filter sets were available and their exact MJDs differed only by a few days. These points are marked in Fig.~\ref{del_mag_fig}. We combined the individual flux errors of the start and end data points to compute the factor-decrease of the flux (see Table~\ref{del_mag}). We also calculated the decrease in magnitudes using the flux-magnitude relation (Appendix~\ref{appendix:abfluxmag}). The largest change is seen in the $c$-band, while the smallest change occurs in the $u'$-band.

%
\begin{table}
\renewcommand{\arraystretch}{1.3}
\caption{Comparison of the outburst strength in three optical filters over a time span of approximately 160 days. The value $\Delta$flux denotes the factor by which the flux decreased over our chosen time-span, while $\Delta{\textrm{mag}}$ gives the value in units of magnitudes. The uncertainties for the flux and magnitude decreases are only based on the statistical errors of the individual data points.}
 \label{del_mag}
\centering                          
\begin{tabular}{c c c c}       
\hline
 & $u'$-band & $c$-band & $o$-band \\ 
\hline
MJD start & 59060.1 & 59058.6 & 59062.6 \\
MJD end & 59219.0 & 59218.3 & 59216.3 \\
$\Delta$flux & $12.1\pm0.1$ & $21.7\pm0.3$ & $19.5\pm0.3$ \\ $\Delta{\textrm{mag}}$ & $2.71\pm0.01$ & $3.34\pm0.02$ & $3.23\pm0.01$ \\
\hline
\end{tabular}
\end{table}


\subsection{Infrared data}

\begin{figure}
   \centering
   \resizebox{\hsize}{!}{\includegraphics{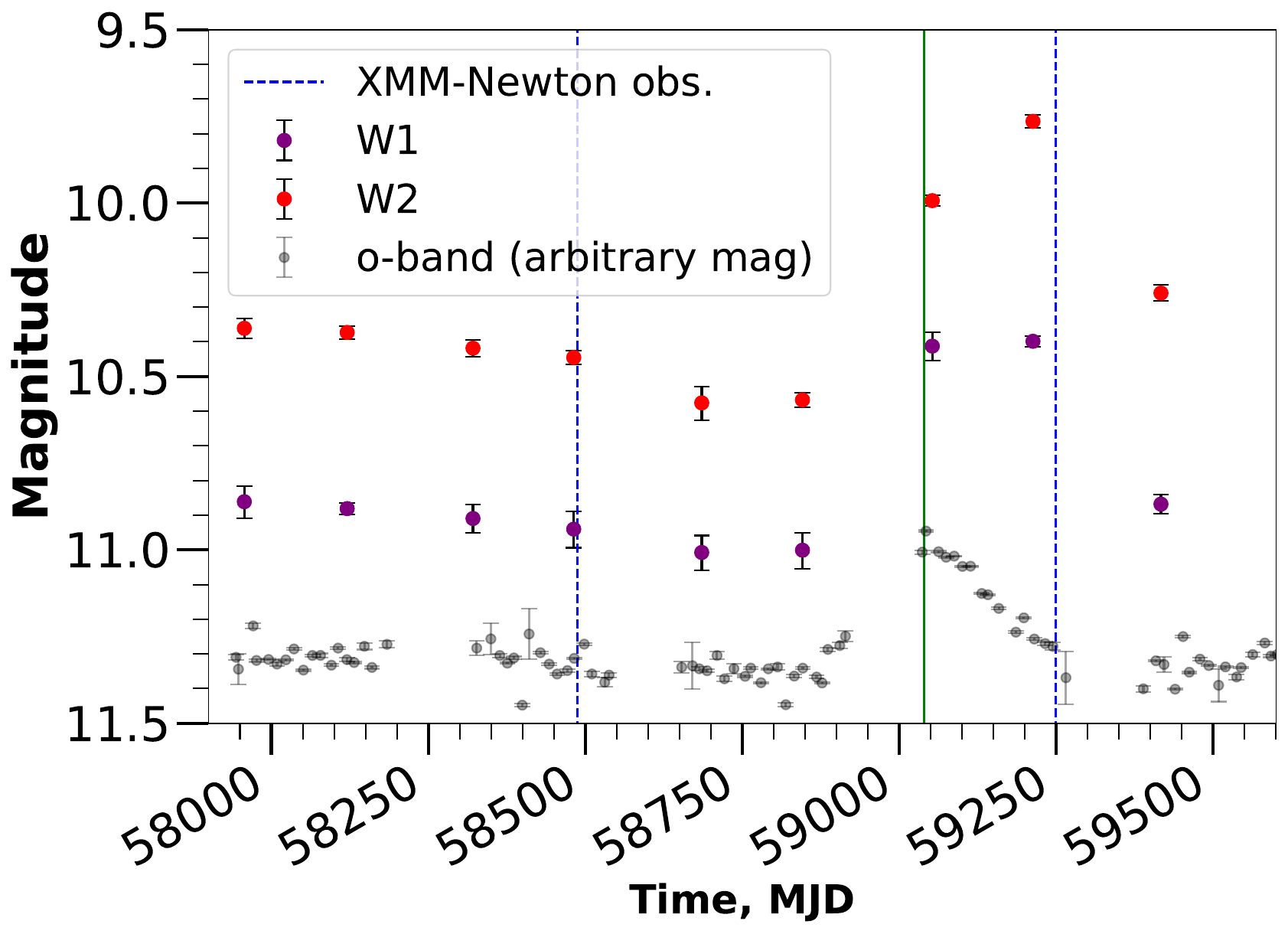}}
   \caption{\textit{WISE} IR light curve from MJD 57957 to MJD 59417 (Jul 2017 -- Jul 2021). The ATLAS $o$-band data for the same time period are shown for a visual reference of the rapidness of the IR response. The green line indicates the observed peak of the optical outburst in the ATLAS $o$-band. The blue dashed lines indicate the dates of the two \textit{XMM-Newton} observations before and after the outburst.}
              \label{wiselc}%
    \end{figure}

We only used \textit {WISE} data from the AGN's faint phase (after 2017; see Fig.~\ref{wiselc}). Unfortunately, precise estimates on the response to the outburst cannot be derived. This is due to the low cadence (one data point every six months) and the rise of the optical outburst being unobservable during sunblock. Nevertheless, it seems to follow the optical outburst fairly quickly if not almost instantaneously. The first W1 and W2 data were only taken 13 days after the $o$-band observed peak in emission output, but the IR data already show a clear increase compared to the data half a year before the 2020 outburst. 

The sparse cadence also does not allow us to determine the precise time when the peak of the IR emission occurred with respect to the optical peak, or if there is a time-delay between W1 and W2. If the W1 and W2 emission peaks occur almost simultaneously, the data distribution suggests an IR-emission peak between MJD 59053 and MJD 59213. Assuming that the IR emission returns to its approximately stable faint-phase flux around the latest data point (MJD 59417), we can say that the outburst lasted between 364 and 570 days in the IR. We note that this is longer than the 200--300 days outburst seen in the optical. In the W1 filter the difference from faintest (last data point before the outburst) to brightest point (MJD 59213) is $\sim$0.6 mags or a factor of $\sim$1.7 and in W2 the observed magnitude difference is $\sim$0.8 mags or a factor of the order of 2.1. 

\subsection{Optical spectra}

The optical spectra are very similar. Our analysis shows no statistical differences between the FWHM and integrated fluxes of both H$\alpha$ and H$\beta$ before and after the outburst. An additional spectrum taken before the outburst covering only H$\beta$ also gives no indication of statistical changes in H$\beta$ compared to the spectrum after the outburst. Our LBT line fitting indicates a narrow H$\beta$ line, which, after correcting for instrumental broadening, shows a dispersion of $\sim$300~\si{\kilo\metre\per\second}. For the FOR2 spectrum, the lines are dominated by the instrumental broadening and, if we correct for this, we find no change within our error margins. We confirm that both OIII line widths are consistent and that the width of H$\beta$ is comparable with these, demonstrating that the H$\beta$ emission we measure in this procedure originates in the NLR.

\subsection{X-ray data}

The first spectrum was taken approximately a year and half before the 2020 outburst was observed in the optical. The second spectrum was closer in time to the peak, being observed approximately seven months after the STELLA light curves shows the brightest magnitude observed during the 2020 outburst. From the model fittings it is evident that the flux has remained or returned to the observed faint state X-ray luminosity of $L_{0.2-10~\textrm{keV}} \sim 7\times10^{42}$ \si{erg.s^{-1}} before and after the outburst. If the X-ray flux scaled with the optical flux during the outburst (flux increase by a factor of the order of 13), the 0.2--10~keV X-ray luminosity could have been as high as  $9.1\times10^{43}$ \si{erg.s^{-1}} during the peak. The most notable difference between the two spectra is seen in the strength of the 6.4 keV Iron emission line. The Fe line has an equivalent width twice as large as in the 2021 spectrum, indicating that changes can be observed even seven months after the optical outburst peak. 

Using the approach from \citet{marconi}, we estimate $L_{\textrm{bol}}/L_{2-10~\textrm{keV}}\sim10$ ($L_{2-10~\textrm{keV}}$ is given in Table~\ref{xray_fit}). We took the mass of the SMBH to be $\log(M_{\textrm{BH}}/M_{\sun})$ = 7.9 (\citealt{mcelroy}; using optical spectroscopic data from 2009). Thus, we estimated the bolometric luminosity before and after the outburst to be $4.8\times10^{43}$~\si{erg.s^{-1}} and the accretion ratio relative to Eddington to be $0.004$.

\subsection{UV data}

Surprisingly, the second UV observation shows an increase of $\sim$0.6 mags compared to before the outburst, even 7--8 months after the observed optical peak. The UV observations were taken simultaneously with the X-ray observations, which do not show an observed increase in the primary X-ray flux after the outburst. 

\section{Discussion} \label{sec5}

\subsection{Multi-wavelength responses to the outburst}
\label{discuss_1}

Compared with the long term light-curve shown in \citet{mcelroy} and \citet{krumpe}, it is evident that the 2020 optical outburst is the most significant as yet observed during Mrk~1018's new type 1.9 state. Unfortunately, our follow-up observations were taken several months after the observed optical peak, due to the large-scale shutdown of many facilities during the 2020 COVID pandemic. We computed the Eddington ratio to be $\sim$0.4\%, based on X-ray observations before and after the outburst. Based on our $u'$-band AGN-only light curve the flux increased by a factor of the order of 13 from the faint state to the observed peak of the outburst. Since the $u'$-band data is also closest in wavelength to the UV emission expected from the accretion disc, we used this factor to estimate the Eddington ratio during the observed peak of the outburst. Thus, we argue that, during the outburst, the AGN was shortly at a value of $\sim$5.2\% of the critical Eddington luminosity limit before returning to 0.4\%.

There is also evidence of an outburst occurring in 2017 \citep{krumpe}. The optical (increase by a factor of 1.6), UV (factor of 1.5) and X-ray flux (factor of 1.9) increase shortly before sunblock in this time period. However, the significance is not known, as the peak and decline are hidden by the sunblock period. We estimated the outburst duration to be a maximum of approximately 210 days so as not to violate the data obtained before and after sunblock in 2017. On the condition that, in 2017, there was an outburst of a similar shape to the 2020 outburst, we speculate that the peak may have been less bright, allowing for a shorter decline. If these outbursts are not merely isolated phenomena, but rather periodic events, our $u'$-band monitoring programme would catch a similar outburst in the future, perhaps in mid-2023.

The high cadence optical ($u'$-, $c$-, and $o$-band), as well as the lower cadence IR monitoring, tracks the decline of the outburst. The optical spectra of the order of eight months before and 17 months after the observed peak of the $u'$-band outburst do not show any differences in the H$\beta$ and H$\alpha$ lines. In particular, no broad line components are detected. We note that \cite{hutsemekers} find faint broad components in their fitting due to their sophisticated reduction technique (e.g. a detailed modelling and removal of the host galaxy contribution). We have used a less complex spectral analysis technique due to our differing goal, namely to compare the BLR before and after the outburst, rather than perform an in-depth spectral analysis. We do not find evidence of changes in the line profiles (e.g. blue wings) possibly indicating outflows as proposed for other CL-AGN such as IRAS 23226-3843 (\citealt{kollatschny_23}). \cite{mcelroy} report broad components in H$\alpha$ and H$\beta$ for both the SDSS (2000) and MUSE (2015) spectra (although the broad H$\beta$ is faint in the MUSE spectrum). The FWHM for H$\beta$ is not reported in the paper, but values of $4000\pm100$~\si{\kilo\metre\per\second} (SDSS) and $3300\pm200$~\si{\kilo\metre\per\second} (MUSE) are given for the broad H$\alpha$ line widths. This is in contrast to our findings of only narrow-line components for H$\beta$ and H$\alpha$ in the more recent spectra. However, we note that the SDSS spectrum is taken in the year 2000, during Mrk~1018's bright type 1 phase, while the 2015 MUSE spectrum is taken during transition. Thus, neither spectra represent the true faint state as reached by the AGN before the outburst in 2020.

In order to gain understanding of how quickly we expect the broad lines to respond, we used two separate methods to calculate the distance of the BLR from the SMBH. No broad line components were needed for the spectral fitting of H$\beta$ and H$\alpha$. Therefore, we cannot use methods relying on broad-line widths or broad-line luminosities.

Firstly, we used the relation from \citet{chen_18}, which relies on the estimated continuum luminosity at 5100~\AA:

 \begin{equation}
     \log \left( \frac{R_{\textrm{BLR}}}{\textrm{light day}} \right) = 
     1.527 + 0.533 
     \log \left( \frac{\lambda L_{\lambda}(5100\, \r{A})}{ 10^{44}\, \si{\erg \per \second} } \right).
 \end{equation}

We calculated $L_{\lambda}(5100 \r{A})$ using the MODS spectrum taken after the outburst, that is the most recent spectrum. However, this includes contribution from both AGN and host and therefore the result of the order of 10 light days may be an overestimation of the BLR radius.

Secondly, we used the relation from \citet{greene_10}, which depends on $L_{2-10~\textrm{keV}}$:

 \begin{equation}
     \log \left ( \frac{R_{\textrm{BLR}}}{10~\textrm{light days}} \right) = 
     0.09 + 0.52
     \log \left( \frac{L_{2-10~\textrm{keV}}}{ 10^{43} \si{erg.s^{-1}} } \right).
 \end{equation}

This method uses the X-ray luminosity as a proxy for AGN luminosity. We estimated a BLR distance of approximately eight light days. Neither of these methods claim to be highly precise, however, they both agree that the BLR is at a distance of the order of ten light days. The fairly short light-travel times give a natural explanation as to why we did not see any changes in the Balmer lines of the order of 17 months after the peak: there was ample time for the BLR to return to its semi-stable low-state configuration. The fact that we saw no changes in the spectrum indicates that we were simply too late to catch any significant response. Nevertheless, this brief increase of the accretion ratio by a factor of approximately 10 does not leave any imprints in the BLR. All physical parameters of the BLR managed to return to the pre-outburst state within 17 months. This also indicates that the outburst is indeed just a transient event.

In the UV, we see an increase in flux by a factor of 1.8 between the \textit{XMM-Newton} observations before and after the outburst. There are two potential explanations for this. Firstly, that this is merely stochastic AGN variability and has no connection to the 2020 outburst. The fact that the X-ray flux did not change between the same \textit{XMM-Newton} observations can be explained as the X-rays are produced by reprocessing (Compton up-scatter) UV/optical photons from the accretion disc. Thus, after the latest \textit{XMM-Newton} observation, there might have also been a (stochastic) X-ray increase. In the second scenario, the UV increase is connected to the 2020 outburst. Since the UV measurement is the combined emission from the host galaxy and AGN components, the flux increase of the AGN itself is likely to be even larger than a factor of 1.8. An additional blackbody (BB) component can explain this UV excess. However, its temperature must be lower than the primary one. Due to the moderate temperature, the second BB would increase the UV emission but have no impact on the combined X-ray flux. Highly speculatively, such a second BB could be produced by a second accretion disc, associated with the second SMBH.

The IR emission is thought to originate from UV/optical photons being reprocessed by the dusty torus and re-emitted in the IR wavelength. Since the reprocessing requires the presence of dust, we calculated the radial distance from the central SMBH where dust sublimates. This is equivalent to the inner radius of the torus. The equation used, as taken from \citet{markowitz_14}, is:

\begin{equation}
    R_{\textrm{d}} \sim 0.4 \left ( \frac{L_{\textrm{bol}}}{10^{45} \si{erg.s^{-1}} } \right )^{1/2} 
    \left( \frac{T_{\textrm{d}}}{T_{1500}} \right )^{-2.6} \si{pc} .
\end{equation}

The parameter $T_{\textrm{d}}$ is the temperature at which dust sublimates, taken to be 1500~\si{K}, as in \citet{nenkova_08}. Inputting the bolometric luminosity for Mrk~1018 gave a dust sublimation radius of about 102 light days. Although Fig.\ref{wiselc} indicates a faster response time, this could be explained by the orientation of the torus. Unless the torus is close to face-on, there would be some amount of dust in our line-of-sight. In that case, we would see the reaction from the area first and fairly quickly after the central flux increase in the UV and optical. That is, the IR re-emission of dust in our direct line-of-sight reaches the telescope more quickly. Photons from the parts of the torus not directly emitting towards us will take much longer, prolonging the IR response. We note that \citet{lamassa} fitted both \textit{Chandra} and \textit{NuSTAR} spectra with a MYTorus model, and found the fit to be consistent with a face-on torus. However, as the inclination is only constrained to be under 60\textdegree, some part of the torus could conceivably intercept our view of the AGN. This line-of-sight theory is a possible explanation for why we see the torus react so quickly.

Firstly, this hypothesis predicts that, due to a combination of light-travel time within the AGN itself and the geometry of the torus, we would observe the likely peak of the IR response within 100 days. Next, radiation emitted directly away from us would have to cross the approximate 100-light-day distance from the accretion disc to the other side of the torus and return, arriving at the telescope about 200 days after the response starts. Lastly, we know that the optical outburst declines in at least 200 days. The fading optical radiation is also re-processed by the torus and the IR decline is extended in time. Adding all these values together gives an IR outburst length of 400--500 days. The observed IR light curve (Fig.~\ref{wiselc}) agrees remarkably well with all these predictions. The increase in IR emission happens very rapidly after the optical peak. Although not known exactly due to the low cadence, a IR peak roughly 100 days after the optical peak is also reasonable. Lastly, the IR outburst duration is much longer than the optical outburst. This scenario predicts that the IR emission will return to its pre-outburst flux level within the next two \textit{WISE} observations.

Traditionally, the torus has been modelled as a smooth doughnut-like shape \citep{orig_torus}, however this model has been refined over the years. Newer models include inhomogeneities (clumps in the form of clouds) in the torus and non-uniform dust temperatures, as the side directly illuminated is hotter \citep{clumpy_torus}. The inner radius of the torus (the dust sublimation radius) could be closer to the central engine as shielding allows some dust to survive here. This configuration significantly cuts the time delay in comparison with a classic smooth torus and isotropically emitting accretion disc, and also explains a quicker than expected IR response. Lastly, in clumpy torus models the clouds are not only located close to the equatorial plane of the AGN \citep{clumpy_torus}. Thus, dusty clouds also exist at very high angles of inclination. In this way, even a torus with a face-on inclination would still show a fast IR response.

In the X-ray observations, the properties of the primary X-ray radiation (power-law components) show no statistically significant difference between the pre-outburst spectrum (approximately 18 months before) and post-outburst spectrum (approximately seven months after). This is plausible as the X-ray response to an outburst in the accretion disc should happen within light-hours to light-days. Thus, the X-ray observation after the outburst was simply too late to catch any immediate responses. The only detected changes in the X-ray spectrum are in the 6.4~keV Fe~line, which is thought to be a reflection feature. If the line is produced at the same distance as the inner region of the torus (as suggested in \citealt{fe_origin}), the time delay to this region will cause a relatively slow response. As a reflection feature, the angle of incidence also plays a part. The parts directly in our line of sight on the near side of torus are less efficient in scattering the radiation towards us, so the Fe~line outburst is not expected to appear quickly after the optical outburst. The Fe~line peak would follow in 100--200 days after the optical peak.

Our \textit{XMM-Newton} observations, obtained only approximately seven months after the probable outburst peak, show the Fe~line with a strength twice as high as before the outburst. Thus, we speculate that the observation was obtained shortly after the largest response of the reflected Fe~line (see also Fig.~\ref{wiselc}, where the time of the second X-ray observation is consistent with being obtained after the IR peak). This scenario suggests that the primary X-ray flux was at least a factor of two higher during the peak of the outburst. A new X-ray observation should show that the Fe line strength has returned by now to its previous flux, as measured in 2019.

\subsection{Potential causes of the outburst}

\citet{mcelroy}, \citet{husemann}, and \citet{krumpe} confirm that the transition from type 1 Seyfert galaxy to type 1.9 Seyfert galaxy almost ten years ago was due to an intrinsic change in accretion flow. This was done by ruling out obscuration, as there was no change seen in the intrinsic absorption of the X-ray spectrum, and a TDE, as the bright period lasted over 30 years -- much longer than can be expected for a TDE. \cite{Noda_2018} suggest that the 2015 changing-look event in the AGN of Mrk~1018 could be triggered by an accretion disc instability mechanism, which propagates throughout the disc and facilitates a state transition. The disagreement with the viscous timescale could be accounted for by domination of either or both radiation and magnetic pressure over gas pressure.

We also see no intrinsic absorption in the X-ray spectra taken before and after the 2020 outburst and thus conclude that the outburst is not caused by a temporary massive decrease of absorbing material along out line of sight. In this scenario the AGN in Mrk~1018 would have emitted on roughly the same level over the last few years. If large amounts of obscuring material blocked the line-of-sight before and after the event, but not during the event, it would have looked like the energy output of Mrk~1018 briefly increased.

Our new multi-wavelength data set indicates that the accretion flow surrounding the SMBH in Mrk~1018 is still varying significantly, even after returning to a type 1.9 Seyfert galaxy. It is unclear whether a similar mechanism to the one that caused the type transition almost a decade ago also caused the intrinsic short-term increase in accretion rate related to the 2020 outburst.

\citet{clumpy_disc} discuss AGN variability in the context of a clumpy accretion disc, rather than a standard, smooth, thin disc model \citep{thin_disc}. They attribute the UV/optical emission to optically thick shocks and the X-ray emission to optically thin shocks within the disc. \citet{disc_models} discuss the fact that some observations show a UV time-lag behind the X-ray emission that suggests a re-processing of X-ray photons by the accretion disc. There is a discrepancy between the disc size inferred from these observations and from the standard thin-disc model. They then explore whether the discrepancy could be explained by a clumpy accretion disc and find that, while this could be a potential cause, it is still unclear whether this is applicable to all AGN. The current literature does not seem to contain an exploration of a clumpy accretion disc in the context of powerful months-long optical outbursts like the one in 2020. Simulations of clumpy disc models are needed to understand if these clumpy structures can survive several orbits in the accretion disc or are smoothed out. If the smoothing does indeed happen relatively quickly, the pronounced, short-term 2020 outburst with a swift return to the previous state would be hard to explain with a clumpy accretion disc.

A warped accretion disc, as outlined in \citet{warped}, is another possible scenario. \citet{disc_tearing} suggest that the warping could cause parts of the disc to break off into discrete rings. Instabilities closer to the central engine could cause short-term quasi-periodic eruptions, whereas those further out could be responsible for variability on changing-look timescales. The disc parameters set the timescale and amplitude of the variability. Relating this to Mrk~1018, such a model may be able to account for both the brief, speculatively periodic, outbursts seen in 2017 and 2020, and the multiple changing-look transitions. However, such models still lack detailed expectations for observations to support or rule out this interesting scenario.

An alternative explanation for short-term outbursts are tidal disruption events (TDEs, \citealt{tde_rees}). There are already several observed cases of this scenario, such as reported in \citet{shaya_19} and \citet{homan_23}. In these scenarios, a rapid increase is followed by a power-law like decline, with a canonical value for the exponent of -5/3 \citep{tde_rees, tde_phinney}. However, \citet{vanvelzen} reports a range of parameters possible for this power-law exponent, with -5/3 being the mean value. In Sects. \ref{lightcurvefitting} and \ref{atlas_fit_sec}, we fitted the $t^{-\frac{5}{3}}$ fixed-exponent power law thought to be characteristic of a TDE. The fitting proved this to be an inappropriate model for the decline in all three optical wavebands. The best-fit function in all three filters is a linear model. Thus, we conclude that the 2020 outburst is not caused by a TDE.

As introduced in Sec.~\ref{intro}, CCA is a key model to interpret extremely rapid changes in AGN light curves. Unlike classical models, CCA is an alternative mechanism to accrete matter onto a SMBH without incorporating the material in a persistent accretion disc. Clouds inelastically colliding head-on, cancelling out angular momentum, and radially driven towards the SMBH, cause a major boost in the intrinsic short-term accretion rate of the AGN (\citealt{gaspari17_micro}), even up to two orders of magnitude in a few years time. Once the cloud is accreted by the SMBH, no further material is available, unlike in an accretion disc scenario. Consequently, the accretion rate drops quickly back to the low-level variability baseline. Such variability boosts or drops can describe the 2020 outburst in Mrk~1018. In this scenario repeated outbursts are possible but they should occur randomly in time.

Another theory for repeated significant AGN variability is that of a BBH at the centre of the AGN \citep{Begelman_1980}. Periodic variability is an important signature of a BBH as the rate of accretion would be altered depending on whether the black holes were at the pericentre or apocentre of their orbits.  If there are regular outbursts -- for example, in 2017, 2020, and 2023 -- we will catch them with our STELLA optical monitoring programme. However, we also note that \cite{hutsemekers} analyse the polarised spectrum of Mrk~1018 and find no evidence of polarisation signatures predicted for a BBH.

Repeated changes to line profiles are further indicators of a BBH. Shifts in emission line profiles or the presence of double-peaked optical and X-ray emission lines, as reported in \citet{severgnini_18}, would indicates the presence of two emitting regions. It is thought that a mini-accretion disc could form around each black hole, with a larger circumbinary disc forming around both \citep{roedig_14}. It is not inconsistent with our data set, but there are no tight constraints that something similar happens with the Fe~line in Mrk~1018. Although in the previous section we discuss the origin of the 6.4~keV Fe line as the inner radius of the torus, it is still unclear which structure predominantly produces the line. It is possible that the line is mainly produced near the outer radius of the accretion disc \citep{fe_origin}. As the black holes orbit each other, the lines from the two emitting regions would be red- or blue-shifted depending on the movement away or towards the observer. If the lines are shifted in opposite directions, depending on resolution, they would appear as one broad line or two separated lines. If we believe that there is a BBH at the centre of Mrk~1018, the broadening of the Fe~line could be due to two unresolved lines shifted in opposite directions. The Fe~line would then broaden and narrow periodically as the black holes orbit around each other. We note that the Fe line is unresolved before the outburst, but the line is resolved afterwards at a confidence level of $\sim3\sigma$. The increase in line width is only in contradiction with the value from before the outburst by $\sim2\sigma$.

Alternatively, there are other processes that can lead to a broadening of the Fe line such as outflowing material at relativistic velocities (e.g. \citealt{mizumoto_18}). \citet{nayakshin_00} introduce another model that is able to explain the change of the Fe line strength. Rather than a continuous and smooth temperature distribution in the disc, their model consists of separate layers with distinct temperatures. Most of the Fe line emission is then produced in the coolest areas. When the layer composition changes (e.g. due to different ionising radiation levels), the Fe line is also expected to change. We also discussed in Sect.~\ref{discuss_1}, the possibility of a secondary, low-temperature BB contributing to the increased UV emission in our second \textit{XMM-Newton} observation. With the caveat that this is pure speculation, this additional BB component could arise from an accretion disc of a second SMBH. The increased UV emission, coupled with periodic outbursts and periodic changes to the Fe~line, would be circumstantial evidence for a BBH scenario.

Lastly, a similar theory of a single recoiling SMBH (rSMBH) in Mrk~1018 was proposed by \citet{Kim_2018}. This is postulated as the reason for the velocity offset between two kinematically shifted components in H$\alpha$ (one red-shifted and one blue-shifted), which is not visible in our data set. Their interpretation of the scenario is two kinematically distinct BLRs, each surviving from the two galaxies that merged to form Mrk~1018. These are gravitationally bound to the rSMBH. They also consider a BBH but discount the idea for the following reasons: the velocity difference between the blue and red line components does not alternate as expected for BBHs; in BBHs the larger SMBH is expected to rotate more slowly but the line widths of the broad-line components show no indications of this; and there are no pairs of narrow emission lines seen. From their simulations \citet{Kim_2018} estimate that the rSMBH has an eccentric orbit of period 29 years. They then suggest the driving force behind Mrk~1018's changing-look behaviour is the tidal impulse that the rSMBH's accretion disc is subject to as it passes the pericentre of its orbit. Because the orbital velocity is larger than the disc rotational velocity, the tidal impulse from the mass of the host galaxy causes density perturbations which affect the accretion rate and, thus, energy output. 

The model by \citet{Kim_2018} describes the extended changing-look behaviour over tens of years, rather than short-term outbursts over months like the 2020 outburst. However, we do not claim that the scenario of a rSMBH is incompatible with smaller outbursts. It could be that two separate mechanisms are responsible for the short- and long-term variable behaviour, or that more detailed simulations will show that the rSMBH model also accounts for smaller outbursts. \citet{Kim_2018} also suggest that the AGN would revert back to a type 1 optical spectrum in the mid-2020's and state that further data are needed to confirm the connection of the broad-line velocity offsets to the recurrent variability in Mrk~1018. Our continued monitoring will provide a light curve that can be compared with the model's predictions.

The 2020 outburst has been caused by a dramatic change in accretion rate but we still do not have a complete picture of why. The proposition of a BBH or rSMBH at the centre of such a clear post-merger remnant is appealing, however, no robust observational evidence for either theory exists yet. Continued $u'$-band monitoring is one key point to check for further outbursts and maybe even periodicity in these events. If a new outburst is found immediate multi-wavelength observations (high cadence X-ray flux, monthly X-ray spectroscopy, monthly UV spectroscopy, monthly optical spectroscopy) have to be available to catch and follow the response of the different AGN structures. Even if no new outburst is found, another deep X-ray spectrum is crucial to confirm or deny our prediction that the Fe~line has returned to its pre-outburst flux.

\section{Conclusions} \label{sec6}

High cadence $u'$-band  monitoring of the AGN in Mrk~1018 caught an impressive outburst in mid-2020. To separate the AGN and host-galaxy contribution in the photometric data, we model the host-galaxy contribution and obtain an AGN-only $u'$-band light curve. Although the rise is blocked by the sun avoidance period, the brightest data point indicates a flux increase by a factor of of the order of 13 compared to the faint phase immediately before. Investigation into the shape of the outburst reveals that i) the outburst is asymmetric with a rise of less than 100 days and a decline of at least 200 days and ii) the best-fit function to the decline is linear. We conclude from the second point that the outburst was not caused by a TDE -- these flares are expected to decline with a power-law shape.

The outburst is also seen by the ATLAS forced photometry server in two redder optical wavelength ranges. These data are also host-subtracted, however the method is not as rigorous as ours. We fitted the decline in the $o$- and $c$-bands and confirmed that the best-fit function is linear. We compared the three data sets by finding data points within about two days of each other in all wavebands and calculating the flux difference. The flux decrease in the $o$- and $c$-bands during the outburst might be higher than in the $u'$-band, but due to the different host galaxy subtraction method the results need to be interpreted with care. 

We then explored the multi-wavelength data available before and after the outburst. As we did not catch the outburst in time due to the unavailability of facilities during the 2020 COVID pandemic, some of the follow-up observations are significantly after the event. Optical spectra were taken approximately eight months before the observed peak in the $u'$-band data and $\sim$17 months after. We compared the H$\beta$ and H$\alpha$ lines in the spectra and find that a single Gaussian is a good fit for both. The FWHM and integrated flux of the two lines agree to within 2$\sigma$ before and after the event. 

Calculations of the BLR radius at approximately ten light days tally with our observations that the region has returned to its previous state within 520 days from the observed optical peak. Nevertheless, it is amazing that no imprint whatsoever is left by this major outburst. No matter what changes did occur, they disappeared so thoroughly that we would have missed the outburst based on optical photometry or spectroscopy with sparse sampling in time. It is possible that similar outbursts may have been missed in other CL-AGN that are not being monitored as frequently as Mrk~1018.

There are simultaneous X-ray and UV \textit{XMM-Newton} observations taken approximately 18 months before and seven months after the observed optical peak. There is no intrinsic absorption seen before and after the outburst, thus ruling out that the increase of emission was due to a temporary decrease in obscuring material. The X-ray flux and X-ray spectral components are also unchanged between observations to a confidence level of $<$2$\sigma$. The X-ray emitting region is extremely close to the central engine, therefore this is not surprising. However, the reflected 6.4 keV Iron line is twice as strong about seven months after the observational outburst peak. This implies that the primary X-ray flux increased by at least a factor of two during the outburst.

The sparse IR light curve already shows a response only around 13 days after the observed optical peak. A potential explanation is that the IR response is able to reach us quickly due to re-processed photons from dust in our line of sight. We calculated the dust sublimation radius, commensurate with the inner radius of the torus, to be approximately 100 light days from the SMBH (in the AGN faint type 1.9 state). A staggered response from regions of the torus at different angles to our line of sight, using our calculated light travel time, agrees with our time-scale estimate.

The multi-wavelength data agree with our calculations of the response time of the various emitting structures of the AGN to the 2020 outburst. The event is most likely caused by a drastic, short-term increase of the accretion rate, for which there are multiple explanations. We discuss several scenarios. Disc models are likely too inefficient to drive the observed rapid increase and decrease in the AGN's light curves. CCA can explain the observed short-term outburst with a few orders of magnitude change in rapid time. Other potential interpretations can be ascribed to a BBH or a rSMBH.

Clearly, the CL-AGN in Mrk~1018 has not finished surprising us. Further high-cadence monitoring, at least in the optical and X-ray, will be crucial to fully unveil Mrk~1018's processes. This new monitoring will test if such major outbursts occur repeatedly, or even periodically. Overall, to answer the ultimate question of why Mrk~1018 repeatedly changes its energy output, multi-wavelength follow-up observations must be immediately be triggered if a new outburst occurs.

\begin{acknowledgements}
R.B. acknowledges that this research was supported by DLR grant number FKZ 50 OR 2004. 
M.K. is supported by DFG grant number KR3338/4-1. 
D.H. acknowledges support from DLR grant FKZ 50 OR 2003.
M.G. acknowledges partial support by HST GO-15890.020/023-A and the {\it BlackHoleWeather} program.\\
This work is based on data obtained with the STELLA robotic telescopes in Tenerife, an AIP facility jointly operated by AIP and IAC. 

This work includes observations obtained with XMM-Newton,
an European Space Agency (ESA) science mission with instruments and contributions directly funded by ESA Member States and NASA.

This work includes observations from FORS2, an instrument on the VLT commissioned by ESO.

This paper uses data taken with the MODS spectrographs built with funding
from NSF grant AST-9987045 and the NSF Telescope System Instrumentation
Program (TSIP), with additional funds from the Ohio Board of Regents and the
Ohio State University Office of Research”.

It has made use of data from the Asteroid Terrestrial-impact Last Alert System (ATLAS) project. The Asteroid Terrestrial-impact Last Alert System (ATLAS) project is primarily funded to search for near earth asteroids through NASA grants NN12AR55G, 80NSSC18K0284, and 80NSSC18K1575; byproducts of the NEO search include images and catalogues from the survey area. This work was partially funded by Kepler/K2 grant J1944/80NSSC19K0112 and HST GO-15889, and STFC grants ST/T000198/1 and ST/S006109/1. The ATLAS science products have been made possible through the contributions of the University of Hawaii Institute for Astronomy, the Queen’s University Belfast, the Space Telescope Science Institute, the South African Astronomical Observatory, and The Millennium Institute of Astrophysics (MAS), Chile.

This publication also makes use of data products from the Wide-field Infrared Survey Explorer, which is a joint project of the University of California, Los Angeles, and the Jet Propulsion Laboratory/California Institute of Technology, funded by the National Aeronautics and Space Administration. 

This research made use of Astropy,\footnote{http://www.astropy.org} a community-developed core Python package for
Astronomy.

\end{acknowledgements}

\bibliographystyle{aa}
\bibliography{mrk1018.bib}

\begin{appendix}

\section{STELLA host galaxy subtraction} \label{appendix:host_subtraction}

\subsubsection{Host galaxy modelling}
\label{hostmodelling}

We used images taken during the faint phase (VIMOS images in the time period 2016--2017) to model the AGN and its host galaxy. Unfortunately STELLA does not have a high enough resolution to visually separate the two components. As a starting point, we used the image obtained on the 14th of February 2018. At that time, the AGN emission was the faintest, making it easier to model the galactic morphology in Mrk~1018.

   \begin{figure}[t]
   \centering
   \resizebox{\hsize}{!}{\includegraphics{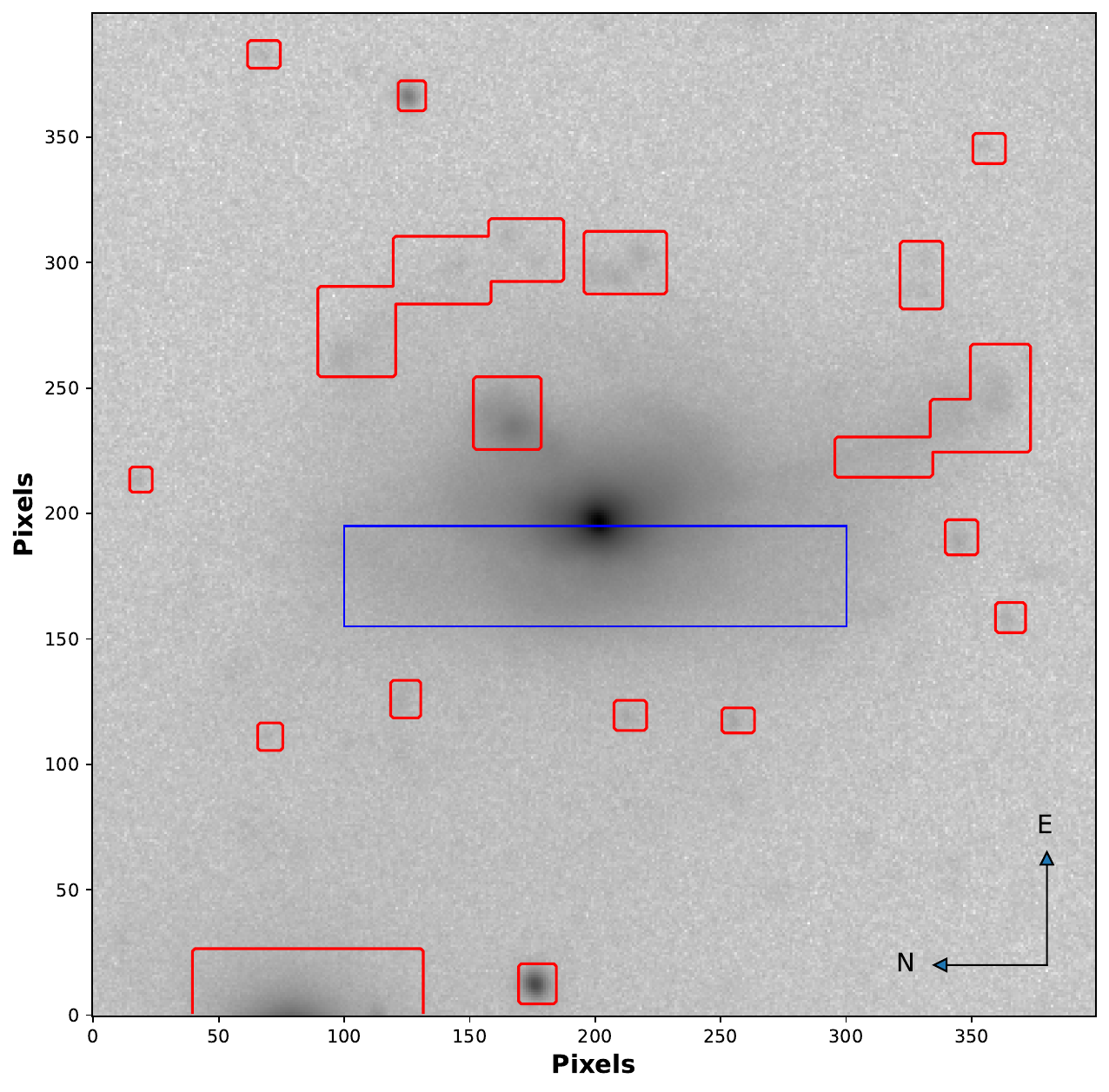}}
      \caption{Image created by stacking the 20 VIMOS exposures of Mrk~1018 during the faint (type~1.9) phase, shown with a log scale. The pixel scale is $0.205\arcsec$, corresponding to 179~\si{pc} per pixel. The red boxes indicate regions that are masked in the fitting process. A subtle light-absorbing structure is marked with a blue box. This is approximated in the fitting by an elongated S\'ersic function with a negative intensity. }
         \label{vimos}
   \end{figure}

We used the image-fitting software Imfit\footnote{https://www.mpe.mpg.de/$\sim$erwin/code/imfit/} \citep{imfit} with a Levenberg-Marquardt algorithm to model Mrk~1018. Our goal was to remove the host galaxy contribution at and around the AGN, rather than creating a highly accurate model over the entire extent of the galaxy. Firstly, we masked the central region containing the AGN ($20\times20$ pixels, or $4.1\arcsec\times4.1\arcsec$) for the initial fitting to determine a best fit of the host galaxy only, which we fitted with a single Sérsic function \citep{sersic}. A Sérsic function is given by the following equation:

\begin{equation} \label{sersic}
    I(R)=I_e \exp\left\{ -b_n\left[ \left( \frac{R}{R_e}\right) ^{1/n} -1\right] \right\}.
\end{equation}

We note that $I_e$ is the half-light intensity, $R_e$ is the half-light radius, $n$ is the Sérsic index (the higher the index, the more centrally concentrated the galaxy's luminosity) and $b_n$ is a constant, dependent on the Sérsic index used \citep{bn}. We also needed to compensate for a large area in the host galaxy affected by dust or some other obscuring material (highlighted in the lower part of Mrk~1018 in Fig.~\ref{vimos}). We therefore added a secondary Sérsic function with a large eccentricity and negative intensity. Further out from the centre, there are additional complex structures in the host galaxy as Mrk~1018 is a post-merger remnant. In order to get Imfit to focus on the central region, rather than trying to fit these irregularities, we created a mask to discount them from the fitting. Figure~\ref{vimos} indicates the regions that we masked for the procedure.

\begin{figure}[t]
   \centering
   \resizebox{\hsize}{!}{\includegraphics{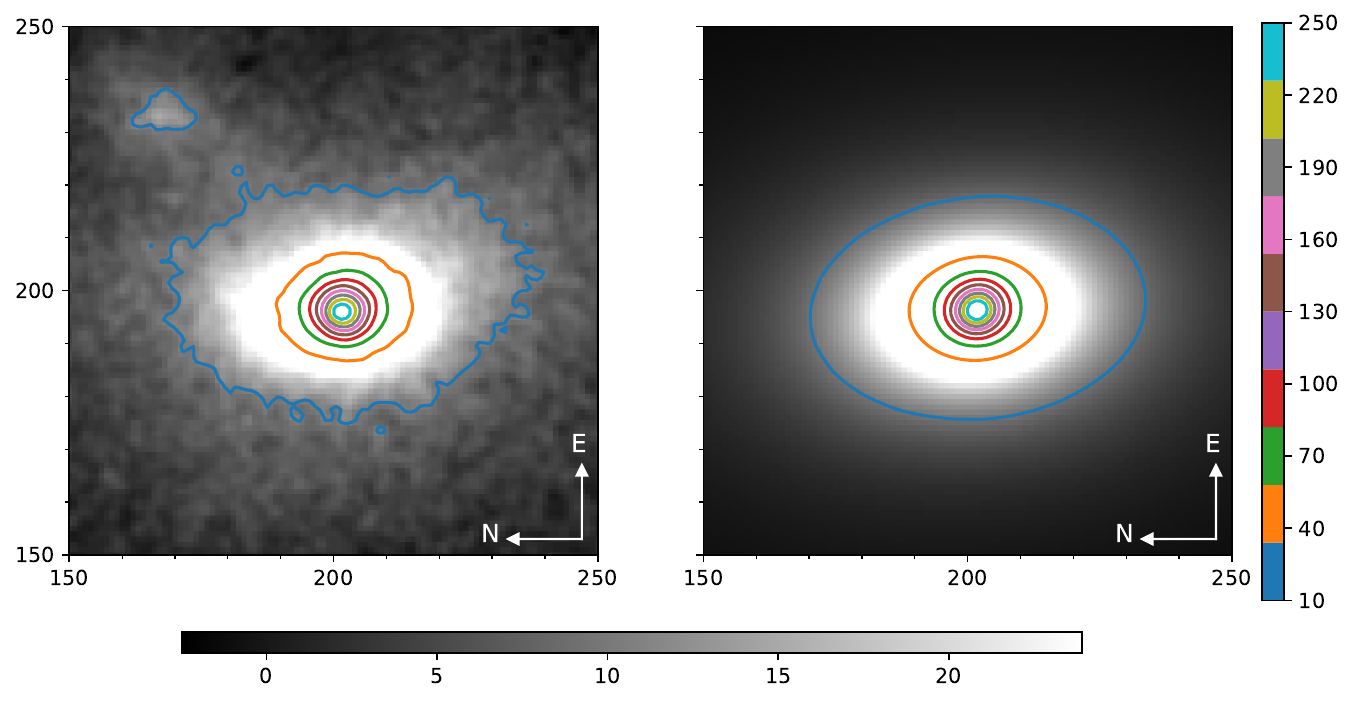}}
      \caption{Observed and modelled images in units of pixels with a spatial resolution of $0.205\arcsec$ per pixel, corresponding to 179~\si{pc} per pixel. The black and white colour bar indicates the number of counts per pixel and the vertical colour bar shows the values of the contour lines. Left: Observed VIMOS host galaxy image after removal of the best-fit (point-like) AGN model. The different coloured lines are contour lines of constant number of counts. This image is used to determine the count contribution of the host galaxy to the total emission of Mrk~1018. Right: Best-fit model of the host galaxy. Comparison of the inner contours in both images shows that the central region is accurately modelled, even if large residuals remain in the outskirts.}
         \label{contourplot}
   \end{figure}

\begin{figure}[t]
   \centering
   \resizebox{\hsize}{!}{\includegraphics{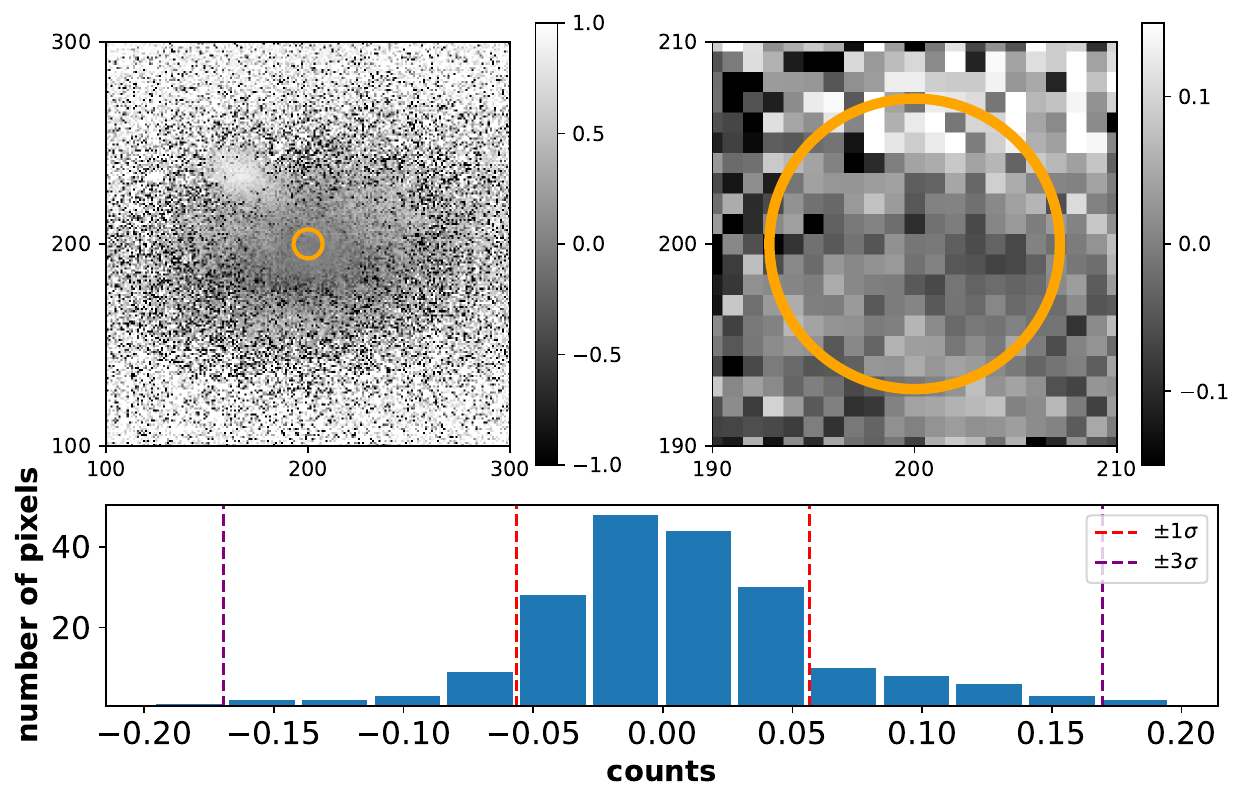}}
   \caption{Top left: Image created by subtracting the Mrk~1018 (AGN and host galaxy) model from the observed VIMOS image then dividing by the observed image. The x- and y-units are pixels and the spatial resolution is $0.205\arcsec$ (179~\si{pc}) per pixel. The orientation is the same as Fig.~\ref{contourplot}. The orange circle encloses 99.7\% (3$\sigma$) of the total point-like PSF (based on reference star three). The values of the pixels show the fractional differences between the model and the science image (from $-$1.0 to $+$1.0, corresponding to a difference of $\pm$100\%). Top right: Image zoomed into the central region. The orange circle, again, indicates the 3$\sigma$ limit of the PSF. The colour scale has been adjusted to a maximum of +0.15 and a minimum of $-$0.15 ($\pm$15\%). Bottom: Distribution of the counts for pixels that fall within the 3$\sigma$ limit of the PSF. The vertical red and purple lines indicate the $\pm1\sigma$ and $\pm3\sigma$ limits of the count distribution respectively. The $1\sigma$ level corresponds to deviations of $\pm$0.06 ($\pm$6\%).}
              \label{modelcheck}%
    \end{figure}

Once we had fitted a model to the host galaxy with Imfit we checked the residuals. Modelling all components precisely in such a complex host galaxy is challenging. Therefore, we aimed for a model that deviated by, on average, less then 10\% from the data in the inner regions of the host galaxy. The AGN-subtracted host galaxy image and best-fit host galaxy model is shown in Fig.~\ref{contourplot}. Once this stipulation was met, we unmasked the AGN and fitted the central area with an additional point source. In practice, this is done by fitting a point spread function (PSF) to the area. Reference star three was inputted to Imfit in order for the algorithm to define the PSF. Once the model (host galaxy plus AGN) was fitted to the data, we checked the overall residuals were within an average deviation of 10\% compared to the observed data (see Fig.~\ref{modelcheck}).

\subsubsection{Measuring the AGN-only contribution}
\label{degrading}

The best-fit Imfit model allowed us to separate the observed counts in Mrk~1018 into two components: the point-like AGN contribution described by a PSF and the extended host galaxy contribution described by two Sérsic functions. Consequently, we could produce a light curve of Mrk~1018's AGN without the host galaxy contribution. In the simplest approach, we used the ratio of the counts from reference star three in the VIMOS image and STELLA images to normalise the VIMOS host galaxy contribution to the STELLA images. However, the STELLA background is much higher than that of the VIMOS background and strongly varies in both space and time. Thus, a significant amount of host galaxy contribution in the STELLA images is lost due to the background. We therefore created a pipeline to approximate the quality of each individual STELLA image in the VIMOS host galaxy image before normalising.

Our pipeline consisted of several steps for each STELLA image per night. Firstly, we re-binned and rotated the VIMOS image to match the pixel scale and orientation of Mrk~1018 in the individual STELLA images (Python package `reproject'\footnote{https://reproject.readthedocs.io/en/stable/}). In this procedure we conserved the integrated flux of Mrk~1018 in the VIMOS observation as measured in our aperture. Secondly, we applied a two dimensional Gaussian filter with a full width half maximum (FWHM) equal to that of the STELLA images\footnote{https://docs.astropy.org/en/stable/api/astropy.convolution.convolve\_ \ fft.html}. This step accounted for the STELLA seeing, which is worse than the VIMOS seeing. Next, we added the STELLA background to the VIMOS-based host-galaxy (only) image. After this we approximated the statistical photon noise associated with STELLA and VIMOS with a random number generator. The steps we took for this process were as follows: we looped through the VIMOS host galaxy image and the STELLA local background image (same dimensions) pixel by pixel; we calculated the Poisson error for each image in each consecutive pixel; we combined these two uncertainties; we generated a number from a normal distribution with a mean of zero and a standard deviation equal to the combined uncertainty; and finally, we added the counts in the VIMOS host image pixel, in the STELLA background pixel, and the number generated by the random number generator and assigned them to the equivalent pixel in a new image. Finally, we subtracted the background from this STELLA-approximated VIMOS host image and measured the counts inside a $10\arcsec$ radius aperture; the same size used in the STELLA data analysis. As VIMOS collects more photons than  STELLA, there were still a significantly higher amount of counts in this image than in a typical STELLA image. As a final step, we therefore normalised VIMOS to STELLA using reference star three.

\subsubsection{Quantifying uncertainties}

We considered two sources of uncertainties in our magnitude calculations of the AGN-only light curve: statistical and systematic. In the above-mentioned pipeline, we used a random number generator when degrading the host galaxy image to STELLA quality. This approximated photon noise. However, this also meant that each run generated a slightly different image. We used this to evaluate the statistical uncertainties. In practice, we ran the pipeline 1000 times for each STELLA image. We then sorted the output array in increasing order and determined the central value, the 159th, and 840th values, respectively. We interpreted these as the median value and its corresponding statistical lower and upper $1\sigma$ bounds.

The systematic uncertainties take the impact of the host galaxy modelling process into account. So far, we only considered the VIMOS image with the faintest AGN contribution for the clearest view of the galaxy's features. However, any image from the VIMOS data set could have been chosen and may have yielded slightly different fitting results. Moreover, we have a set of high quality images from the Gemini Multi-Object Spectrograph South (GMOS-S) which, although not as high-resolution as the VIMOS images, can also be used for modelling. To quantify this systematic uncertainty on the host galaxy modelling, we chose eight VIMOS and two GMOS images from the set of 20 VIMOS images and 21 GMOS images. We chose them to span the observing time from August 2016 to January 2019.  We then repeated all the steps summarised under Sects.~\ref{hostmodelling} \&~\ref{degrading} for each of these images. Next, we did photometry with a $10\arcsec$ radius aperture on the central region of the (AGN-subtracted, STELLA-quality) host galaxy images. This resulted in a range of values for the host galaxy counts. We then calculated the mean and standard deviation of this range. We selected three images from the range: one with the host galaxy count value closest to the mean of the distribution (referred to as 'mean image'), and the other two with host galaxy count values closest to the $\pm$1$\sigma$ standard deviation values. The mean image was our final choice for our host galaxy image and was used to calculate the final light curve data points and their statistical uncertainties as described above (see Fig.~\ref{stella_lc}).

We used the other two host galaxy images (with counts closest to the $\pm$1$\sigma$ values) to calculate the systematic uncertainties, shown by the shaded region in Fig. \ref{stella_lc}. For each of these two selected host galaxy images, we ran the pipeline 1000 times, only taking the central values of the two outputted distributions, and derived the corresponding upper and low systematic uncertainties of the light curve. We ignored the 159th and 840th values in these distributions as we are not interested in the statistical uncertainties here. The resultant light curve is shown in Fig.~\ref{stella_light_curve}.

\newpage

\section{Extended ATLAS light curve} \label{appendix:big_atlas_lc}

   \begin{figure}[h]
   \centering
   \resizebox{\hsize}{!}{\includegraphics{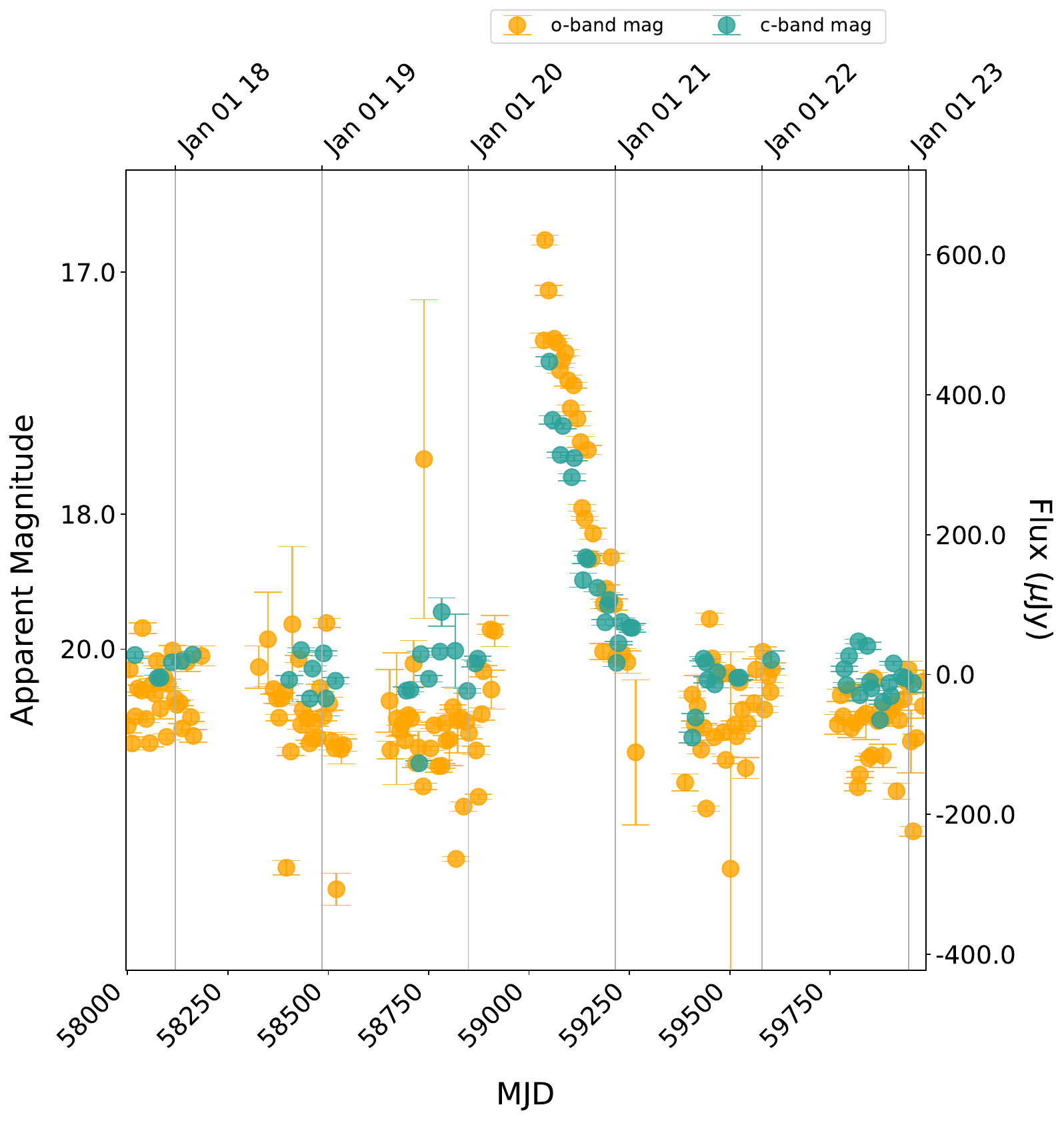}}
   \caption{Optical light curve downloaded from the ATLAS forced photometry server spanning several years before and after the 2020 outburst. The data were binned to a 7-day cadence. We note that these data show the difference fluxes and magnitudes compared to an ATLAS reference image. The plot indicates that several years before and after the 2020 outburst the AGN remained in a semi-stable faint state.}
    \end{figure}
%

\section{AB flux-magnitude relation} \label{appendix:abfluxmag}

AB magnitudes, $m_{AB}$, can be converted to AB fluxes, $f_{AB}$, in Janskys as outlined in \citet{fluxmag}. The equation used is:

\begin{equation}
    f_{AB} = 10^{- \frac{m_{AB}}{2.5}} \times 3631~\mathrm{Jy}.
\end{equation}

The corresponding uncertainties in the flux can be found from the magnitude uncertainties \citep{erprop} using this relation, where $\sigma_f$ and $\sigma_m$ are the flux and magnitude relations respectively:

\begin{equation}
    \sigma_f = \left| f_{AB} \times\frac{\ln(10)}{-2.5} \sigma_m \right| .
\end{equation}

\end{appendix}

\end{document}